\documentclass[titlepage,12pt]{utarticle}
\usepackage{amsmath,amssymb,amsthm,amscd,cite}
\usepackage{latexsym,graphicx}
\usepackage[all]{xy}
\xyoption{v2}
\xyoption{2cell}
\xyoption{dvips}
\CompileMatrices
\LaTeXdiagrams
\UseAllTwocells
\newdimen\tableauside\tableauside=1.0ex
\newdimen\tableaurule\tableaurule=0.4pt
\newdimen\tableaustep
\def\phantomhrule#1{\hbox{\vbox to0pt{\hrule height\tableaurule width#1\vss}}}
\def\phantomvrule#1{\vbox{\hbox to0pt{\vrule width\tableaurule height#1\hss}}}
\def\sqr{\vbox{%
  \phantomhrule\tableaustep
  \hbox{\phantomvrule\tableaustep\kern\tableaustep\phantomvrule\tableaustep}%
  \hbox{\vbox{\phantomhrule\tableauside}\kern-\tableaurule}}}
\def\squares#1{\hbox{\count0=#1\noindent\loop\sqr
  \advance\count0 by-1 \ifnum\count0>0\repeat}}
\def\tableau#1{\vcenter{\offinterlineskip
  \tableaustep=\tableauside\advance\tableaustep by-\tableaurule
  \kern\normallineskip\hbox
    {\kern\normallineskip\vbox
      {\gettableau#1 0 }%
     \kern\normallineskip\kern\tableaurule}%
  \kern\normallineskip\kern\tableaurule}}
\def\gettableau#1 {\ifnum#1=0\let\next=\null\else
  \squares{#1}\let\next=\gettableau\fi\next}
\numberwithin{equation}{section}

\newcommand{\be}{\begin{equation}}
\newcommand{\ee}{\end{equation}}

\newcommand\IZ{\mathbb {Z}}
\newcommand{\IC}{\mathbb{C}}

\newcommand{\IR}{\mathbb{R}}
\newcommand{\ba}{\begin{array}}
\newcommand{\ea}{\end{array}}

\newcommand{\CV}{{\mathcal V}}

\newcommand{\CK}{{\cal K}}
\newcommand{\kah}{{K\"ahler}}

\newcommand{\IH}{{\mathbb H}} 
\newcommand{\bal}{\begin{aligned}}
\newcommand{\eal}{\end{aligned}}
\newcommand{\half}{{1\over 2}}
\newcommand{\re}{{\hbox{Re}}}
\newcommand{\im}{{\hbox{Im}}}
\newcommand{\ch}{{\hbox{ch}}}
\newcommand{\td}{{\hbox{Td}}}
\begin{document}
\preprint{
    {\tt hep-th/0602138}
}
\title{A D-Brane Landscape on Calabi-Yau Manifolds}
\author{Duiliu-Emanuel Diaconescu\footnote{\tt duiliu@physics.rutgers.edu},\ \ 
Alberto Garcia-Raboso\footnote{\tt agraboso@physics.rutgers.edu}\\
and Kuver Sinha\footnote{\tt kuver@physics.rutgers.edu}}
\oneaddress{
      {\centerline  {\it Department of Physics and Astronomy, 
Rutgers University,}}
      \smallskip
      {\centerline {\it Piscataway, NJ 08854-0849, USA}}
      }
\date{}

\Abstract{
We explore the dynamics of magnetized nonsupersymmetric D5-brane 
configurations on Calabi-Yau orientifolds with fluxes. We show that
supergravity D-terms capture supersymmetry breaking effects predicted by 
more abstract $\Pi$-stability considerations. We also examine 
superpotential interactions in the presence of fluxes, and investigate the 
vacuum structure of such configurations. Based on the shape of the potential, 
we argue that metastable nonsupersymmetric vacua can be in principle obtained 
by tuning the values of fluxes.} 

\maketitle

\section{Introduction}

Magnetized branes in toroidal IIB orientifolds have been a very useful device 
in the construction of semirealistic string vacua 
\cite{Blumenhagen:2003vr,Cascales:2003pt,Cascales:2003zp,Larosa:2003mz,Cremades:2004wa,Font:2004cx,
Cvetic:2004xx,Lust:2004dn,Marchesano:2004xz,Marchesano:2004yq,Cvetic:2005bn}.
A very attractive feature of magnetized brane systems is \kah\ moduli
stabilization by D-term effects 
\cite{Burgess:2003ic,Antoniadis:2004pp,Kors:2004hz,Antoniadis:2005nu,Dudas:2005pr,Dudas:2005vv,
GarciadelMoral:2005js}. 
By turning on background fluxes, one can stabilize the complex structure 
moduli as well, obtaining an interesting distribution of isolated 
vacua in the string theory landscape. These are typically supersymmetric 
vacua because magnetized brane configurations are supersymmetric for special 
values of the toroidal moduli. Note however, that nonsupersymmetric vacua 
have also been found in \cite{Kors:2004hz,Dudas:2005pr,Dudas:2005vv} as a result of 
the interaction between D-term and nonperturbative F-term effects. 

The purpose of the present work is to explore the landscape of magnetized 
brane configurations on Calabi-Yau manifolds. The starting point of this 
investigation is the observation that certain Calabi-Yau orientifolds
exhibit a very interesting class of metastable D-brane configurations. 
As opposed to toroidal models, these brane configurations are not
supersymmetric for any values of the moduli, but the supersymmetry breaking 
parameter is minimal at the Landau-Ginzburg point in the underlying 
$N=2$ moduli space. 
In this paper we investigate the dynamics of these brane configurations from the 
point of view of the low energy effective supergravity action. We compute the 
D-term contribution to the potential energy and show that it agrees with 
more abstract $\Pi$-stability considerations. A similar relation between 
supergravity D-terms and the perturbative part of $\Pi$-stability was 
previously found in \cite{Blumenhagen:2005pm}.
We also develop a generalization
of the flux superpotential in the presence of magnetized branes. 
Then we argue that the interplay between D-term effects and the flux 
superpotential can in principle give rise to a landscape of 
metastable nonsupersymmetric vacua. Note that different aspects of the
open string landscape have been recently studied in 
\cite{Blumenhagen:2004xx,Gmeiner:2005vz,Gmeiner:2005nh,Gomis:2005wc}.   

Let us briefly outline of our construction. We will consider IIB
orientifolds of Calabi-Yau manifolds with $h^{1,1}=1$ which have only
space-filling O3 planes. 
Our main example, described in detail in
section two, is an orientifold of the octic hypersurface in weighted
projective space $WP^{1,1,1,1,4}$. 
The D-brane configuration consists of a D5-brane 
wrapping a holomorphic curve $C$ and an anti-D5-brane wrapping 
the image curve $C'$ under the orientifold projection. Both $C,C'$ are 
rigid and do not intersect each other. We also turn on worldvolume $U(1)$ 
magnetic fluxes so that each brane has $p$ units of induced D3-brane charge. 

Such configurations are obviously nonsupersymmetric, at least for generic 
values of the \kah\ moduli, since D5-branes and O3 planes do not preserve the 
same fraction of supersymmetry. The supersymmetry breaking parameter 
can be taken to be the phase difference between the central charges of these 
objects in the underlying $N=2$ theory. This phase can be computed using 
standard $\Pi$-stability techniques, and depends on the complexified \kah\ moduli
of the $N=2$ theory. We will perform detailed computations for the octic 
orientifold example in section three and appendix A. The outcome of these 
computations is that this system is not supersymmetric anywhere on the 
real subspace of the $N=2$ \kah\ moduli space preserved by the orientifold 
projection. However the supersymmetry breaking parameter reaches a minimum 
at the Landau-Ginzburg point. This is a new dynamical aspect which has 
not been encountered before in toroidal orientifolds.

In flat space we would expect this system to decay to a 
supersymmetric configuration of space-filling D3-branes. 
The dynamics is different on Calabi-Yau manifolds since the curves $C,C'$ are rigid, 
which means that the branes have no moduli. This can be viewed as a potential 
barrier in configuration space opposing brane anti-brane annihilation. 
If the branes are sufficiently far apart, so that the open string spectrum does not 
contain tachyons and the attractive force is weak, we will obtain a metastable 
configuration. The system can still decay, but the decay has to be 
realized by tunelling effects. 

This construction already poses a problem since the $N=1$ dynamics is very 
hard to control in a nongeometric phase of the \kah\ moduli space. Ideally one 
would like to describe the theory in terms of a large volume compactification 
so that the $\alpha'$ corrections are small. This can be achieved in the
present context using orientifold mirror symmetry
\cite{Acharya:2002ag,Brunner:2003zm,Brunner:2004zd,Grimm:2004ua}. 
Since the supersymmetry breaking phase is independent on complex structure 
moduli, we can take the IIB Calabi-Yau manifold to be near the 
large complex structure limit point. In this regime, the theory has an a
alternative description in terms of a large volume IIA compactification, 
which will allow us to control the dynamics. Taking this limit, 
we will be able to compute the D-term effects in section three. 
We will also show that the results agree with the $\Pi$-stability analysis.

Moduli stabilization in this system can be achieved by turning on IIA 
fluxes as in
\cite{Behrndt:2004km,Derendinger:2004jn,Kachru:2004jr,Camara:2005dc,Camara:2005pr,
DeWolfe:2005uu,Grimm:2004ua,House:2005yc,Villadoro:2005cu}
Since we also have branes in the picture, 
it turns out that the most convenient description of the flux superpotential 
involves a combination of IIB and IIA variables. This is a special case of the 
bi-period superpotentials introduced in \cite{Berglund:2005dm}, except that we have 
to take into account the D-brane superpotential as well. 
The F-term effects in the presence of branes and fluxes are described in
section four, together with some general aspects of the D-brane configuration
space. Our discussion of the brane-flux superpotential builds on previous 
work on this subject
\cite{Donaldson:1996kp,Witten:1997ep,Lerche:2002yw,Lerche:2002ck,clemens-2002-,Diaconescu:2005jv},
emphasizing the relation between the geometry 
and the light open-string spectrum.  


Finally, in section five we investigate the vacuum structure of the D-brane 
landscape. We analyze the shape of the potential energy, and formulate 
sufficient conditions for the existence of nonsupersymmetric metastable
vacua. Then we argue that these conditions can be in principle satisfied 
by tuning the values of background fluxes. In principle this mechanism can 
give rise to either de Sitter or anti de Sitter vacua, providing an
alternative to the existing constructions of de Sitter vacua 
\cite{Kachru:2003aw,Burgess:2003ic,Escoda:2003fa,Saltman:2004sn,
Saltman:2004jh,Becker:2004gw,Buchbinder:2004im,Saueressig:2005es}
in string theory. 

{\it Acknowledgments.} We would like to thank Bobby Acharya, Frederik Denef, 
Mike Douglas, Bogdan Florea, Robert Karp and Greg Moore for very helpful conversations. 
We owe special thanks to Bogdan Florea and Robert Karp for help with the results 
of appendix A. 

{\it Note added.} When this paper was ready for submission, two new papers 
appeared \cite{Villadoro:2006ia,Martucci:2006ij} which have partial
overlap with our D-term and F-term computations in sections 3 and 4.

\section{A Mirror Pair of Calabi-Yau Orientifolds} 

In this section we review some general aspects of Calabi-Yau orientifolds 
and present our main example. We will first describe the model in IIB 
variables and then use mirror symmetry to write down the low energy 
effective action in a specific region in parameter space. 

Let us consider a $N=2$ IIB compactification on a Calabi-Yau manifold X. Such 
compactifications have a moduli space 
$\CM_h\times \CM_v$ of exactly flat directions, where $\CM_h$ denotes the
hypermultiplet moduli space 
and $\CM_v$ denotes the vector multiplet moduli space. It is a standard fact
that $\CM_h$ must be quaternionic 
manifold whereas $\CM_v$ must be a special K\"ahler manifold. 
The dilaton field is a  hypermultiplet
component, therefore the geometry of $\CM_h$ receives both $\alpha'$ and $g_s$ corrections. 
By contrast, the geometry of $\CM_v$ is exact at tree level in both $\alpha'$ and $g_s$.  
The hypermultiplet moduli space $\CM_h$ contains a subspace $\CM_h^0$ parameterized by vacuum 
expectation values of NS-NS fields, the RR moduli being set to zero. At string tree level 
$\CM_h^0$ has a special \kah\ structure which  
receives nonperturbative $\alpha'$ corrections. These corrections can be exactly summed using 
mirror symmetry. 

Given a $N=2$ compactification, we construct a $N=1$ theory by gauging a discrete symmetry of
the form $(-1)^{\epsilon F_L}\Omega\sigma$ where $\Omega$ denotes world-sheet parity,
$F_L$ is left-moving fermion number and $\epsilon$ takes values $0,1$ depending on the model. 
$\sigma: X \to X$ is a holomorphic involution of $X$ preserving the 
holomorphic three-form $\Omega_X$ up to sign $$\sigma^* \Omega_X = (-1)^\epsilon \Omega_X.$$
We will take $\epsilon=1$, which corresponds to theories with O3/O7 planes. 
In order to keep the technical complications to a minimum, in this paper 
we will focus on models with $h^{1,1}=1$ which exhibit only O3 planes. 
More general models could be treated in principle along the same lines, 
but the details would be more involved. 

According to \cite{Grimm:2004uq}, the massless spectrum of $N=1$ orientifold compactifications 
can be organized in vector and chiral multiplets. For orientifolds with O3/O7
planes, there are $h^{2,1}_-$ chiral multiplets corresponding to
invariant complex structure deformations of $X$, $h^{1,1}_+$ chiral multiplets 
corresponding to 
invariant complexified K\"ahler deformations of $X$, and $h^{1,1}_-$
chiral multiplets parameterizing 
the expectation values of the two-form fields $(B,C^{(2)})$. Moreover, 
we have a dilaton-axion modulus $\tau$. Note that the real K\"ahler deformations 
of $X$ are paired up with expectation values of the four-form field $C^{(4)}$
giving rise to the 
$h^{1,1}_+$ complexified K\"ahler moduli. 
Note also that for one parameter models i.e. $h^{1,1}=1$, we have 
$h^{1,1}_-=0$, hence there are no theta angles $(B,C^{(2)})$. 

The moduli space of the $N=1$ theory must be a \kah\ manifold. 
For small string coupling and large compactification radius the moduli 
space is a direct product between complex structure moduli, complexified 
\kah\ moduli and a dilaton-axion factor. 
The \kah\ geometry of the moduli space can be determined in this regime by 
KK reduction of ten dimensional supergravity \cite{Grimm:2004uq}.

For more general values of parameters, the geometry receives both $\alpha'$
and $g_s$ corrections which may not preserve the direct product structure. 
In particular, we expect significant $\alpha'$ corrections in nongeometric 
regions of the \kah\ moduli space such as the Landau-Ginzburg phase. 
There is however a different regime in which the geometry of the  
moduli space is under control, although the \kah\ parameters take 
nongeometric values. This follows from mirror symmetry for 
orientifolds \cite{Acharya:2002ag,Brunner:2003zm,Brunner:2004zd,Grimm:2004ua}.

Mirror symmetry relates the IIB $N=2$ compactification on $X$ to a IIA 
$N=2$ compactification on the mirror Calabi-Yau manifold $Y$. 
The complex structure moduli space $\CM_v$ of $X$ is identified to the \kah\ 
moduli space of $Y$. In particular, there is a special boundary point of 
$\CM_v$ -- the large complex structure limit point (LCS) -- which is 
mapped to the large radius limit point of $Y$. Therefore if the complex 
structure of the IIB threefold $X$ is close to LCS point, we can find an 
alternative description of a large radius IIA compactification on $Y$. 
This is valid for any values of the \kah\ parameters of $X$, including 
the region centered around the LG point, which is mapped to the LG
point in the complex structure moduli space of $Y$. 

Orientifold models follow the same pattern. Orientifold mirror symmetry relates 
a Calabi-Yau threefold $(X,\sigma)$ with holomorphic involution 
to a threefold $(Y,\eta)$ equipped with an antiholomorphic involution 
$\eta$. As long as the holomorphic involution preserves the large 
complex limit of $X$, we can map the theory to a large radius 
IIA orientifold on $Y$ which admits a supergravity description. 
At the same time, we can take the \kah\ parameters of $X$ close to the LG point, which
is mapped to the LG point in the complex structure moduli space of $Y$. 
This is the regime we will be mostly interested in throughout this paper.

In this limit, the moduli space of the theory has a direct product structure 
\cite{Grimm:2004ua}
\be\label{eq:modspace}
\CM \times \CK
\ee
where $\CM$ is the complex structure moduli space of the IIB orientifold 
$(X,\sigma)$ and $\CK$ parameterizes the complex structure moduli space of the IIA 
orientifold $(Y, \eta)$ and the dilaton. $\CM$ can also be identified with 
the \kah\ moduli space of the IIA orientifold, but the description in terms 
of IIB variables will be more convenient for our purposes.  
We discuss a specific example in more detail below.  

\subsection{Orientifolds of Octic Hypersurfaces}

Our example consists of degree eight hypersurfaces in the
weighted projective space $WP^{1,1,1,1,4}$. The defining equation of an octic
hypersurface $X$ is   
\be\label{eq:octicA}
P(x_1,\ldots, x_5) =0 
\ee
where $P$ is a homogeneous polynomial of degree eight with respect to the $\IC^*$ action 
\[ 
(x_1,x_2,x_3,x_4,x_5) \to (\lambda x_1,\lambda x_2, \lambda x_3, \lambda x_4, \lambda^4 x_5).\] 
This is a one-parameter model with $h^{1,1}(X)=1$ and $h^{2,1}(X)=149$.
 
In order to construct an orientifold model, consider a family of such
hypersurfaces of the form 
\be\label{eq:octicB} 
Q(x_1,\ldots, x_4) + x_5(x_5+\mu x_1x_2x_3x_4) =0  
\ee
where $Q(x_1,\ldots,x_4)$ is a degree eight homogeneous polynomial, 
and $\mu$ is a complex parameter. We will denote these hypersurfaces by 
$X_{Q,\mu}$. 
Consider also a family of holomorphic involutions of $WP^{1,1,1,1,4}$
of the form 
\be\label{eq:holinvA} 
\sigma_\mu : (x_1,x_2,x_3,x_4,x_5)\to(-x_3,-x_4,-x_1,-x_2,-x_5-\mu x_1x_2x_3x_4)
\ee
Note that a hypersurface $X_{Q,\mu}$ is invariant under the holomorphic
involution $\sigma_\mu$ if and only if $Q$ is invariant under the involution 
\be\label{eq:holinvB} 
(x_1,x_2,x_3,x_4) \to (-x_3,-x_4,-x_1,-x_2).
\ee 
We will take the moduli space $\CM$ to be the moduli space of hypersurfaces 
$X_{Q,\mu}$ with $Q$ invariant under \eqref{eq:holinvB}. A similar involution 
has been considered in a different context in \cite{Giryavets:2003vd}.  

One can easily check that the restriction of $\sigma_\mu$ to any invariant 
hypersurface $X_{Q,\mu}$ has finitely many fixed points on $X_{Q,\mu}$ 
with homogeneous coordinates 
\[
\left(x_1,x_2,\pm x_1, \pm x_2, -{\mu\over 2}x_1x_2x_3x_4\right)
\]
where $(x_1,x_2)$ satisfy 
\[
Q(x_1,x_2,\pm x_1, \pm x_2) -{\mu^2\over 4} x_1^4 x_2^4=0.
\] 
Moreover the LCS limit point $\mu \to \infty$ is obviously a boundary point 
of $\CM$. This will serve as a concrete example throughout this paper. 

Mirror symmetry identifies the complexified \kah\ moduli space 
$\CM_h^0$ of the underlying $N=2$ theory to the complex
structure moduli space of the family of mirror hypersurfaces $Y$
\be\label{eq:mirrorA} 
x_1^8+x_2^8+x_3^8+x_4^8+x_5^2-\alpha x_1 x_2 x_3 x_4 x_5 =0 
\ee 
in $WP^{1,1,1,1,4}/\left(\IZ_8^2\times \IZ_2\right)$
\cite{Font:1992uk,Klemm:1992tx,Berglund:1993ax}.
At the same time the complex structure moduli space $\CM_v$ of octic hypersurfaces 
is isomorphic to the complexified \kah\ moduli space of $Y$. 
Orientifold mirror symmetry relates the IIB orientifold $(X,\sigma)$ to a IIA orientifold 
determined by $(Y,\eta)$ where $\eta$ is a antiholomorphic involution 
of $Y$. 

For future reference, let us provide some details on the \kah\ geometry 
of the moduli space following \cite{Grimm:2004ua}. Let $z^i$, 
$i=1,\ldots, h^{1,2}_-(X)$, be algebraic algebraic coordinates on 
the complex structure moduli space $\CM$. The \kah\ potential 
for $\CM$ in a neighborhood of the large complex structure is given by
\be\label{eq:kahlerA} 
K_\CM = -\hbox{ln}\left(i \int_X \Omega_X\wedge {\overline \Omega}_X\right)
\ee 
where $\Omega_X$ is the global holomorphic three-form on $X$. 
This expression is naturally a function  of algebraic coordinates on the IIB 
complex structure moduli space. If we express it in terms of special 
coordinates adapted to the LCS limit, we will   
obtain the tree level \kah\ potential for the IIA \kah\ moduli 
space \cite{Grimm:2004ua} plus $\alpha'$ 
corrections which are exponentially small near the large radius 
limit. 

The second factor $\CK$ parameterizes complex structure moduli of IIA
orientifold and the dilaton. The corresponding moduli fields are \cite{Grimm:2004ua} 
the real complex parameters of $Y$ and the periods of three-form 
RR potential $C^{(3)}$ preserved by the antiholomorphic
involution plus the IIA dilaton.

The antiholomorphic involution preserves 
the real subspace $\alpha = {\overline \alpha}$ of the 
$N=2$ moduli space. 
This follows from the fact that the IIB B-field
is projected out using the mirror map 
\[ 
B+iJ = {1\over 2\pi i} \hbox{ln}(z) + \ldots
\]
where $z=\alpha^{-8}$ is the natural coordinate on the moduli 
space of hypersurfaces \eqref{eq:mirrorA} near the LCS point. 

According to \cite{Grimm:2004ua} (section 3.3), the \kah\ geometry of $\CK$ can be 
described in terms of periods of the three-form $\Omega_Y$ and 
the flat RR three-form $C_3$ on cycles in $Y$ on a   
symplectic basis of invariant or anti-invariant three-cycles on $Y$ 
with respect to the antiholomorphic involution.  
We will choose a symplectic basis of invariant cycles $(\alpha_0,\alpha_1;
\beta^0,\beta^1)$ adapted to the large complex limit $\alpha \to \infty$ of the  
family \eqref{eq:mirrorA}. Using standard mirror symmetry technology,  
one can compute the corresponding period vector $(Z^0,Z^1; \CF_0, \CF_1)$
near the large complex structure limit by solving the Picard-Fuchs equation. 
Our notation is so that the asymptotic behavior of the periods 
as $\alpha \to \infty$ is 
\[ 
Z^0 \sim 1 \qquad Z^1 \sim \hbox{ln}(z) \qquad 
\CF_1 \sim (\ln(z))^2 \qquad \CF_0\sim (\ln(z))^3. 
\]
Moreover, we also have the following reality conditions on the 
real axis $\alpha \in \IR$ 
\be\label{eq:realcondA} 
\hbox{Im}(Z^0) = \hbox{Im}(\CF_1) =0 \qquad 
\hbox{Re}(Z^1) = \hbox{Re}(\CF_0) =0. 
\ee
This reflects the fact that $(\alpha_0,\beta^1)$ are invariant and 
$(\alpha^1,\beta_0)$ are anti-invariant under the holomorphic involution. 
The exact expressions of these periods can be found in appendix A. 
Note that the reality conditions \eqref{eq:realcondA} are an incarnation 
of the orientifold constraints (3.45) of \cite{Grimm:2004ua} in our model. 
In particular, the compensator field $C$ defined in \cite{Grimm:2004ua} 
is real in our case, i.e. the phase $e^{-i\theta}$ introduced in \cite{Grimm:2004ua}
equals 1. 

The holomorphic coordinates on the moduli space $\CK$ are 
\be\label{eq:holcoordA} 
\bal
& \tau = \half \xi^0 + i C \hbox{Re}(Z^0) \cr
& \rho = i {\widetilde \xi}_1 -2 C \hbox{Re}(\CF_1) \cr
\eal
\ee
where $(\xi^0,{\widetilde \xi}_1)$ 
are the periods of the three-form field $C^{(3)}$ on the invariant
three-cycles $(\alpha_0,\beta^1)$
\be\label{eq:Cfieldperiods} 
C^{(3)}= \xi^0\alpha_0 - {\widetilde \xi}_1 \beta^1.  
\ee

Mirror symmetry identifies $(\tau, \rho)$ with the IIB dilaton 
and respectively orientifold complexified \kah\ parameter \cite{Grimm:2004ua}, 
section 6.2.1. A priori, $(\tau,\rho)$ are defined in a neighborhood 
of the LCS, but they can be analytically continued to other 
regions of the moduli space. We will be interested in neighborhood 
of the Landau-Ginzburg point $\alpha =0$, where there is a 
natural basis of periods $[w_2\ w_1\ w_0\ w_7]^{tr}$ 
constructed in \cite{Klemm:1992tx}. The notation and explicit expressions for 
these periods are reviewed in appendix A. 
For future reference, note that the LCS periods $(Z^0,\CF_1)$ in equation 
\eqref{eq:holcoordA} are related to the LG periods by 
\be\label{eq:transmapA} 
\left[\begin{array}{c}
Z^0 \cr Z^1 \cr \CF_1 \cr \CF_0 
\end{array} \right] = 
\left[\begin{array}{cccc} 
0 & 0 & 1 & 0 \cr
\half & \half & -\half & -\half \cr
-{1\over 2} & -{3\over 2} & -{3\over 2} & -{1\over 2} \cr
-1 & 1 & 0 & 0 \cr
\end{array} 
\right]
\left[
\begin{array}{c}
w_2 \cr w_1 \cr w_0 \cr w_7 \cr
\end{array} 
\right]
\ee
Note that this basis is not identical to the symplectic basis of periods 
computed in \cite{Klemm:1992tx}; the later does not obey the reality 
conditions \eqref{eq:realcondA} so we had to perform a symplectic change of
basis. 

The compensator field $C$ is given by 
\be\label{eq:compensator}
C = e^{-\Phi}e^{K_0(\alpha)/2}
\ee
where $e^{\Phi}=e^\phi \hbox{vol}(Y)^{-1/2}$ is the four dimensional IIA dilaton, and 
\be\label{eq:kahlerB}
\bal
K_0(\alpha) & = -\ln\left(i \int_{Y} \Omega_{Y} 
\wedge {\overline \Omega}_{Y} \right)\bigg |_{\alpha={\overline \alpha}}\cr
& = -\ln \left[2\left(\im(Z^1)\re(\CF_1)-\re(Z^0)\im(\CF_0)\right)\right]\cr
\eal
\ee
is the \kah\ potential of the $N=2$ complex structure moduli 
space of $Y$ restricted to the real subspace $\alpha={\overline \alpha}$. 
The \kah\ potential of the orientifold moduli space is given by \cite{Grimm:2004ua}
\be\label{eq:kahlerC}
\bal
K_\CK & = -2 \ln\left(2\int_Y \re(C\Omega_Y) \wedge \ast \re(C\Omega_Y)\right)\cr
& = -2\ln \left[2C^2\left(\im(Z^1)\re(\CF_1)-\re(Z^0)\im(\CF_0)\right)\right]. \cr
\eal
\ee
Note that equations  \eqref{eq:holcoordA}, \eqref{eq:compensator} define 
$K_\CK$ implicitly as a function of $(\tau,\rho)$. The \kah\ potential \eqref{eq:kahlerC} 
can also be written as 
\be\label{eq:kahlerD}
K_\CK= -\ln(e^{-4\Phi})  
\ee
where $\Phi$ is the four dimensional dilaton. 
Let us conclude this section with a discussion of superpotential
interactions. 

\subsection{Superpotential Interactions}

There are several types of superpotential interactions in this system, depending
on the types of background fluxes. Since the theory has a large 
radius IIA description, an obvious option is turning on even RR fluxes
$F^A=F_2+F_4+F_6$ as well as NS-NS flux $H^A$ on the manifold $Y$. 
In principle one can also turn on the zero-form flux $F_0$ as in 
\cite{DeWolfe:2005uu,House:2005yc}, 
but we will set to zero throughout this paper. 

Even RR fluxes give rise to a superpotential for type IIA \kah\ moduli 
of the form \cite{Gukov:1999ya,Gukov:1999gr,Grimm:2004ua,Kachru:2004jr}
\be\label{eq:IIAsuperA} 
W^A_\CM = \int_Y F^A \wedge e^{-J_Y}, 
\ee
where $J_Y$ is the \kah\ form of $Y$. 
The type IIA NS-NS flux is odd under the orientifold
projection, therefore it will have an expansion 
\be\label{eq:IIAfluxA}
H^A = q_1\alpha^1 - p^0 \beta_0. 
\ee 
According to \cite{Grimm:2004ua}, this yields a superpotential for the IIA 
complex structure moduli of the form 
\be\label{eq:IIAsuperB} 
W^A_\CK = -2p^0\tau -i q_1 \rho.  
\ee 

The superpotential \eqref{eq:IIAsuperA} can be given a IIB interpretation 
using mirror symmetry. Recall that in large volume IIB compactifications,
one usually has a flux induced superpotential \cite{Gukov:1999ya} 
\be\label{eq:IIBsuperA} 
W^B = \int_{X} \Omega_X \wedge F^B
\ee 
where $F^B$ is the three-form RR flux on $X$. 
For a comprehensive review of IIB flux compactifications with a 
complete list of references see \cite{Grana:2005jc}. 
Based on the nonrenormalization result of 
\cite{Burgess:2005jx}, this superpotential does not receive perturbative 
$\alpha'$ or $g_s$ corrections. Therefore this superpotential formula 
should still be valid for small values of the IIB \kah\ modulus, although 
we may not have a clear microscopic description of the fluxes. 
Then the superpotential \eqref{eq:IIAsuperA} can be regarded as a IIB 
superpotential of the form \eqref{eq:IIBsuperA}, where 
$F^B$ is the IIB RR flux related by mirror symmetry to $F^A$. 
Using the mirror map, one can show that the two expressions 
agree near the LCS point of the IIB moduli space up to exponentially small 
corrections. For us, it will be more convenient to use the IIB expression, 
keeping in mind that this is just a reformulation of the large radius 
IIA superpotential.

In principle, one could also turn the IIB NS-NS flux $H^B$, but the IIA 
description of the theory would be more involved. According to 
\cite{Gurrieri:2002wz}, the mirror type IIA theory would be a compactification 
on a manifold with a half-flat $SU(3)$ structure. We will not review this 
conjecture in detail here. It suffices to note that granting this conjecture 
one can reformulate the IIB superpotential 
\[
-\int_X \Omega_X \wedge \tau H^B 
\]
in IIA variables \cite{Gurrieri:2002wz}. More details can be found in 
\cite{Camara:2005dc,Camara:2005pr,House:2005yc}. 
In this paper we will not turn on IIB NS-NS flux, but it may be helpful 
to keep in mind that we also have this option. 

In conclusion, in the absence of branes, we will have a total superpotential of the form 
\be\label{eq:totsuper}
W = W^B + W_\CK^A.  
\ee
This formula has to be modified in the presence of magnetized branes. We will
discuss the necessary modifications in section 4. 

We would like to conclude this section with a remark about tadpole 
cancellation. Since we have set the IIB NS-NS flux $H^B$ and the 
type IIA zero-form flux $F_0$ to zero, the only sources for RR tadpoles 
are the orientifold planes and the background D-branes. Magnetized 
D5-branes can also contribute to the tadpole because they carry 
induced D3-brane charge. Therefore the tadpole cancellation condition 
can be written as 
\be\label{eq:tadpole}
 N_{D3} + N_{O3}+p = 0,
\ee
where $p$ is the induced D3-brane charge of magnetized D5-branes.
As explained in the next section, the best option for us is to saturate 
this condition by taking $N_{D3}=0$, i.e. no background D3-branes. 
Let us turn now to magnetized brane configurations.

\section{Magnetized Branes on Calabi-Yau Orientifolds} 

In this section we study the dynamics of magnetized D5-branes wrapping 
holomorphic curves in Calabi-Yau threefolds. We will analyze their 
dynamics both from the world-sheet and low energy supergravity point of view. 
The world-sheet analysis is based on $\Pi$-stability considerations in the 
underlying $N=2$ theory \cite{Douglas:2000ah,Douglas:2000gi,Aspinwall:2001dz}.
Using mirror symmetry, we will show that the world-sheet aspects are captured 
by D-term effects in the IIA supergravity effective action. 
Similar computations have been performed for Type I D9-branes in \cite{Blumenhagen:2005pm}, 
for IIB D3 and D7-branes on Calabi-Yau orientifolds in 
\cite{Grana:2003ek,Lust:2004fi,Lust:2005bd,Jockers:2004yj,Jockers:2005zy}, 
and for D6-branes in toroidal models in  
\cite{Burgess:2003ic,Antoniadis:2004pp,Kors:2004hz,Antoniadis:2005nu,Dudas:2005pr,Dudas:2005vv,
GarciadelMoral:2005js}.
In particular, a relation between the perturbative part of $\Pi$-stability 
($\mu$-stability) and supergravity D-terms has been found in \cite{Blumenhagen:2005pm}.
D6-brane configurations in toroidal models have been thoroughly analyzed from 
the world-sheet point of view in \cite{Lust:2004cx,Bertolini:2005qh}. 
Earlier work 
on the subject in the context of rigid supersymmetric theories includes 
\cite{Douglas:1996sw,Kachru:1999vj,Harvey:2001wm,Lawrence:2004sm}. Our setup is in fact 
very similar to the situation analyzed in \cite{Kachru:1999vj}, except that we 
perform a systematic supergravity analysis. Finally, a conjectural formula
for the D-term potential energy on D6-branes has been proposed in 
\cite{Blumenhagen:2002wn,Lust:2004ks} based on general supersymmetry 
arguments. We will explain the relation between their expression and the 
supergravity computation at the end of section 3.2. Let us start with the 
$\Pi$-stability analysis. 

\subsection{$\Pi$-stability and magnetized D-branes} 

From the world-sheet point of view, a wrapped D5-brane is described by a boundary 
conformal field theory which is a product between an internal CFT factor 
and a flat space factor. 
Aspects related to $\Pi$-stability and superpotential deformations depend only
on the internal CFT part
and are independent on the rank of the brane in the uncompactified four
dimensions. For example the 
same considerations apply equally well to a IIB D5-brane wrapping $C$ or to a
IIA D2-brane wrapping the same curve. 
The difference between these two cases resides in the manner of describing the
dynamics of the 
lightest modes in terms of an effective action on the uncompactified directions of the brane. 
Since the D5-brane is space filling the effective action has to be written in
terms of four dimensional 
supergravity as opposed to the D2-brane effective action, which reduces to
quantum mechanics. 
Nevertheless we would like to stress that in both cases the open string
spectrum and the dynamics 
of the system is determined by identical internal CFT theories; only the low
energy effective 
description 
of these effects is different. Keeping this point in mind, in this section we
proceed with the analysis of the internal CFT factor. 

Although our arguments are fairly general, 
for concreteness we will focus on the octic hypersurface in $WP^{1,1,1,1,4}$.
Other models can be easily treated along the same lines.  
Suppose we have a D5-brane wrapping a degree one rational curve $C\subset X$. 
Note that curvature effects induce one unit of spacefilling 
D3-brane charge as shown in appendix A. 
In order to obtain a pure D5-brane state we have to turn on a compensating
magnetic flux in the  
$U(1)$ Chan-Paton bundle 
$$ 
{1\over 2\pi} \int_C F = -1. 
$$ 
However for our purposes we need to consider states with higher D3-charge,
therefore we will turn on $(p-1)$ units of magnetic flux 
$$ 
{1\over 2\pi} \int_C F = p-1
$$ 
obtaining a total effective $D3$ charge equal to $p$.  
The orientifold projection will map this brane to a anti-brane wrapping $C'=\sigma(C)$ 
with $(-p-1)$ units of flux, where the shift by $2$ units is again  
a curvature effect computed in appendix A.

We will first focus on the underlying $N=2$ theory.  Note that this  
system breaks tree level supersymmetry because the brane and the anti-brane 
preserve different fractions of the bulk $N=2$ supersymmetry. The $N=1$
supersymmetry preserved by a brane is determined by its 
central charge which is a function of the complexified \kah\ moduli. 
The central charges of our objects are 
\be\label{eq:centralchA} 
Z_+ = Z_{D5} + pZ_{D3} \qquad Z_- = -Z_{D5} +pZ_{D3}
\ee
where the label $\pm$ refers to the brane and the anti-brane respectively.  
$Z_{D5}$ is the central charge  
of a pure D5-brane state, and $Z_{D3}$ is the central charge of a D3-brane on
$X$.
The phases of $Z_+,Z_-$ are not aligned for generic values of the \kah\
parameters, but they
will be aligned along a marginal stability locus where $Z_{D5}=0$. 
If this locus is nonempty, these two objects preserve identical fractions of
supersymmetry, and 
their low energy dynamics can be described by a supersymmetric gauge theory. 
If we deform the bulk \kah\ structure away from the $Z_{D5}=0$ locus, 
we expect the brane world-volume 
supersymmetry to be broken. Ignoring supergravity effects, this supersymmetry 
breaking can be modeled by Fayet-Iliopoulos couplings in the 
low energy gauge theory. We will provide a supergravity 
description of the dynamics in the next subsection. 
This effective description is valid at weak string coupling and 
in a small neighborhood of the marginal stability locus in the \kah\ 
moduli space. For large deformations away from this locus the effective gauge
theory description breaks 
down, and we would have to employ string field theory for an accurate description of D-brane 
dynamics. 

Returning to the orientifold model, note that the orientifold projection leaves invariant 
only a real dimensional subspace of the $N=2$ \kah\ moduli space, because it projects out 
the NS-NS $B$-field. As explained in section 2.1, the IIB complexified \kah\
moduli space can be identified with the complex structure moduli space of the family of mirror 
hypersurfaces \eqref{eq:mirrorA}. The subspace left invariant by the
orientifold projection is $\alpha ={\overline \alpha}$. 

Therefore it suffices to analyze 
the D-brane system along this real subspace of the moduli space. 
Note that orientifold $O3$ planes preserve the same 
fraction of supersymmetry as D3-branes. Therefore the above $D5-{\overline {D5}}$ 
configuration would still be supersymmetric along the locus $Z_{D5}=0$ because 
the central charges \eqref{eq:centralchA} are aligned with $Z_{D3}$.
Analogous brane configurations have been considered in \cite{Diaconescu:2005pc} for 
F-theory compactifications. 
  
A bulk \kah\ deformation away from the supersymmetric locus will couple to the 
world-volume theory as a D-term because this is a disc effect which does not change 
in the presence of the orientifold projection. This will be an accurate description 
of the system as long as the string coupling is sufficiently small and we can 
ignore higher order effects. Note that the $Z_{D5}=0$ locus will generically intersect  
the real subspace of the moduli space along a finite (possibly empty) set. 

To summarize the above discussion, the dynamics of the brane 
anti-brane system in the $N=1$ orientifold model can be captured by 
D-term effects at weak string coupling 
and in a small neighborhood of the marginal stability locus 
$Z_{D5}=0$ in the \kah\ moduli space. 
Therefore our first concern should be to find the intersection between the 
marginal stability locus and the real subspace $\alpha ={\overline \alpha}$ 
of the moduli space. A standard computation performed in appendix A shows that the central 
charges $Z_{D3},Z_{D5}$ are given by 
\[ 
Z_{D3} = Z^0\qquad Z_{D_5} = Z^1. 
\]
in terms of the periods $(Z^0,Z^1;\CF^1,\CF^0)$ introduced in section 2.1.
Then the formulas \eqref{eq:centralchA} become 
\be\label{eq:centralchB}
Z_{+}=pZ^0+Z^1,\qquad Z_{-}=pZ^0-Z^1.
\ee
In appendix A we show that the relative phase 
\be\label{eq:relphaseA} 
\theta = {1\over \pi} \left(\im\ln(Z_+) - \im \ln(Z_{D3})\right) 
\ee
between $Z_+$ and $Z_{D3}$ does not vanish anywhere on the real axis 
$\alpha={\overline \alpha}$ and has a minimum at the Landau-Ginzburg point
$\alpha=0$. The value of $\theta$ at the minimum is approximatively $\theta_{min} \sim
1/p$. For illustration, we represent in fig 1. the dependence
$\theta=\theta(\alpha)$ near the Landau-Ginzburg 
point for three different values of $p$, $p=10,20,30$. Note that the minimum value of 
theta is $\theta_{min}\sim 0.12$, therefore we expect the dynamics to 
have a low energy supergravity description. 

\begin{figure}[ht]
\begin{center}
\includegraphics[scale=0.5]{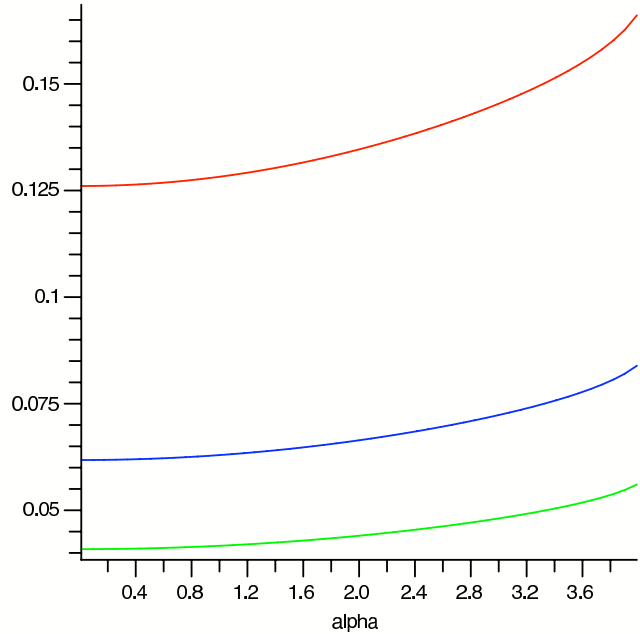} 
\end{center}
\caption{The behavior of the relative phase $\theta$ near the LG point for 
three different values of $p$. Red corresponds to $p=10$, blue corresponds 
to $p=20$ and green corresponds to $p=30$.}
\label{shapeA}
\end{figure}

It is clear from this discussion that the best option for us is to take 
the number $p$ as high as possible subject to the tadpole cancellation 
constraints \eqref{eq:tadpole}. This implies that there are no background 
D3-branes in the system, and we set $p=N_{O3}$. In fact configurations 
with background D3-branes would not be stable since there would be an 
attractive force between D3-branes and magnetized D5-branes. Therefore 
the system will naturally decay to a configuration in which all D3-branes 
have been converted into magnetic flux on D5-branes.  

In order for the above construction to be 
valid, one has to check whether the D3-brane and D5-brane are stable BPS 
states at the Landau-Ginzburg point. This is clear in a neighborhood of the 
large radius limit, but in principle, these BPS states could decay before we 
reach the Landau-Ginzburg point. 
For example it is known that in the $\IC^2/\IZ_3$ local model the D5-brane 
decays before we reach the orbifold point in the \kah\ moduli space
\cite{Douglas:2000qw}. The behavior of the BPS spectrum of compact Calabi-Yau threefolds 
is less understood at the present stage. At best one can check stability 
of a BPS state with respect to a particular decay channel
employing $\Pi$-stability techniques \cite{Aspinwall:2001dz,Douglas:2000ah,Douglas:2000gi}, 
but we cannot rigorously prove stability using the formalism developed in 
\cite{Tom:2002,Tom:2005}. In appendix A we show that 
magnetized D5-branes on the octic are stable with respect to the most natural decay 
channels as we approach the Landau-Ginzburg point. This is compelling evidence  
for their stability in this region of the moduli space, but not a rigorous 
proof. 
Based on this amount of evidence, we will assume in the following 
that these D-branes are stable in a neighborhood of the Landau-Ginzburg point. 
Our next task is the computation of supergravity D-terms in the mirror 
IIA orientifold described in section 2.1. 
 
\subsection{Mirror Symmetry and Supergravity D-terms} 

The above $\Pi$-stability arguments are independent of complex 
structure deformations of the IIB threefold $X$. We can exploit this 
feature to our advantage by working in a neighborhood of the 
LCS point in the complex structure moduli space of $X$. In this region, 
the theory admits an alternative description as a large volume 
IIA orientifold on the mirror threefold $Y$. The details have been 
discussed in section 2.1. In the following we will use the IIA 
description in order to compute the D-term effects on magnetized 
branes. 

Open string mirror symmetry maps the D5-branes wrapping $C,C'$ to
D6-branes wrapping special lagrangian cycles $M,M'$ in $Y$. 
Since $C,C'$ are rigid disjoint $(-1,-1)$ curves for generic moduli of $X$,  
$M,M'$ must be rigid disjoint three-spheres in $Y$. 
The calibration conditions for $M,M'$ are of the form 
\be\label{eq:calib}
\im(e^{i\theta}\Omega_Y|_M)=0\qquad 
\im(e^{-i\theta}\Omega_Y|_{M'})=0. 
\ee
where $\Omega_Y$ is normalized so that the calibration of the 
IIA orientifold O6-planes has phase $1$. 
The phase $e^{i\theta}$ in \eqref{eq:calib} is equal to the 
relative phase \eqref{eq:relphaseA} computed above, 
and depends only on the complex structure moduli of $Y$.
The homology classes of these cycles can be read off from the central charge 
formula \eqref{eq:centralchB}. We have
\be\label{eq:branecharges}
[M]= p\beta^0+\beta^1,\qquad 
[M']= p\beta^0 -\beta^1
\ee
where $[M],[M']$ are  cohomology 
classes on $Y$ related to $M,M'$ by Poincar\'e duality.

Taking into account $N=1$ supergravity constraints, the D-term contribution 
is of the form 
\be\label{eq:DtermA} 
U_D={D^2\over 2 \im(g)}
\ee 
where $g$ is the holomorphic coupling constant of the brane $U(1)$ vector
multiplet.
The holomorphic coupling constant can be easily determined by identifying 
the four dimensional axion field $a$ which has a coupling of the form 
\be\label{eq:axion}
\int a F \wedge F 
\ee
with the $U(1)$ gauge field on the brane. Such couplings are obtained
by dimensional reduction of Chern-Simons terms of the form  
action. 
\[
\int C^{(3)}\wedge F\wedge F + C^{(5)} \wedge F 
\]
in the D6-brane world-volume action.
Taking into account the expression \eqref{eq:Cfieldperiods} for 
$C^{(3)}$, dimensional reduction of the Chern-Simons term on the 
cycle $M$ yields the following four-dimensional couplings 
\be\label{eq:fourdcouplings} 
p\int \xi^0 F\wedge F + \int D^1 \wedge F. 
\ee
Here $\xi^0$ is the axion defined in \eqref{eq:Cfieldperiods} and $D^1$ 
is the two-form field obtained by reduction of $C^{(5)}$ 
\[ 
C^{(5)} = D^1\wedge \alpha^1.
\]
Equation \eqref{eq:fourdcouplings} shows that the axion field $a$ 
in \eqref{eq:axion} is $\xi^0$. Then, using holomorphy and equation
\eqref{eq:holcoordA}, it follows that the tree level
holomorphic gauge coupling $g$ must be  
\be\label{eq:gaugecoupling} 
g = 2p \tau. 
\ee

The second coupling in \eqref{eq:fourdcouplings} is also very useful. The
two-form field $D^1$ is part of an $N=1$ linear multiplet $L^1$ whose 
lowest component is the real field $e^{2\Phi} \im(Z^1)$, where $\Phi$ is the 
four dimensional dilaton \cite{Grimm:2004ua}. Moreover, one can relate $L$ to the chiral 
multiplet $\rho$ by a duality transformation which converts the second 
term in \eqref{eq:fourdcouplings} into a coupling of the form 
\[
\int A_\mu \partial^\mu {\widetilde \xi}_1. 
\]
The supersymmetric completion of this term determines the supergravity D-term 
to be \cite{Dine:1987xk,Dine:1987gj,Binetruy:2004hh,Freedman:2005up}
\be\label{eq:DtermB} 
D =\partial_\rho K_\CK.  
\ee
Note that using  equation (B.9) in \cite{Grimm:2004ua}, the D-term \eqref{eq:DtermB} 
can be written as 
\be\label{eq:DtermBB}
D=-2 e^{2\Phi} \im(CZ^1)
\ee 
where $C$ is the compensator field defined in equation
\eqref{eq:compensator}. 
Using equations \eqref{eq:holcoordA} and \eqref{eq:kahlerD}, we can rewrite 
\eqref{eq:DtermBB} as 
\be\label{eq:DtermBC} 
\bal
D & = -2 e^{K_\CK/2} \im(CZ^1)\cr
& = -{1\over C} {\im(Z^1) \over \im(Z^1)\re(\CF_1)-\re(Z^0)\im(\CF_0)}\cr
& = -{1\over \im(\tau)}{\re(Z^0)\im(Z^1) \over \im(Z^1)\re(\CF_1)
-\re(Z^0)\im(\CF_0)}.\cr
\eal 
\ee 
Then, taking into account \eqref{eq:gaugecoupling}, we find the following
expression for the D-term potential energy 
\be\label{eq:Denergy} 
U_D = {1\over 4p \im(\tau)^3}\left[ 
{\re(Z^0)\im(Z^1) \over \im(Z^1)\re(\CF_1)-\re(Z^0)\im(\CF_0)}\right]^2.
\ee
This is our final formula for the D-term potential energy. 

In order to conclude this section, we would like to explain the relation 
between formula \eqref{eq:Denergy} and the $\Pi$-stability analysis
performed earlier in this section.  
Note that the $\Pi$-stability considerations are captured by an effective 
potential in the mirror type IIA theory which was found in 
\cite{Blumenhagen:2002wn,Lust:2004ks}. 
According to \cite{Blumenhagen:2002wn,Lust:2004ks}, 
the D-term potential for a pair of D6-branes as above should be given by 
\be\label{eq:DtermC} 
V_D = 2e^{-\Phi} \left(\bigg|\int_M {\widehat \Omega}_Y\bigg| - \int_M 
\re({\widehat \Omega}_Y)\right)
\ee
where ${\widehat \Omega}_Y$ is the holomorphic three-form on $Y$ 
normalized so that 
\[
i \int_Y {\widehat \Omega}_Y \wedge {\overline {\widehat \Omega}_Y} = 1.
\]
Recall that $\Phi$ denotes the four dimensional dilaton. 

In the following we would like to explain that this expression is in agreement 
with the supergravity formula \eqref{eq:Denergy} for a small supersymmetry 
breaking angle $|\theta|<<1$. For large $|\theta|$ the effective supergravity 
description of the theory breaks down, and we would have to employ string 
field theory in order to obtain reliable results. 

Note that one can write 
\be\label{eq:normform}
{\widehat \Omega}_Y = e^{K_0/2}\Omega_Y
\ee
where $K_0$ is has been defined in equation \eqref{eq:kahlerB}, and $\Omega_Y$ 
has some arbitrary normalization.
The expression in the right hand side of this equation is left invariant under 
rescaling $\Omega_Y$ by a nonzero constant. 

Formula \eqref{eq:DtermC} is written in the string frame. In order to compare
it with the supergravity expression, we have to rewrite it in the Einstein 
frame. In the present context, the string metric has to be rescaled by a
factor of $e^{2\phi}(\hbox{vol}(Y))^{-1}=e^{2\Phi}$ \cite{Kors:2003wf}, 
hence the potential energy in the Einstein frame is 
\be\label{eq:DtermD}
V_D^E=2e^{3\Phi}\left(\bigg|\int_M {\widehat \Omega}_Y\bigg| - \int_M 
\re({\widehat \Omega}_Y)\right).
\ee
Taking into account equations \eqref{eq:branecharges} and \eqref{eq:normform} we have 
\[
\bal 
\int_M {\widehat \Omega}_Y & = e^{K_0/2}(p\re(Z^0)+i\im(Z^1)) =
e^{K_0/2} Z_{+} 
\eal
\]
where $Z_+$ is the central charge defined in equation \eqref{eq:centralchA}. 
For small values of the phase, $|\theta|<<1$, we can expand \eqref{eq:DtermD} as 
\be\label{eq:DtermE} 
V_D^E\sim e^{3\Phi}e^{K_0/2}{\re(Z_0)\over p} \left[{\im(Z^1)\over
\re(Z^0)}\right]^2. 
\ee
Now, using equations \eqref{eq:holcoordA} and \eqref{eq:DtermBB} in
\eqref{eq:DtermA}, we obtain 
\be\label{eq:DtermF}
\bal 
U_D & = C e^{4\Phi}{\re(Z^0)\over p}\left[{\im(Z^1)\over \re(Z^0)}\right]^2
= e^{3\Phi}e^{K_0/2} {\re(Z^0)\over p}\left[{\im(Z^1)\over \re(Z^0)}\right]^2\cr
\eal 
\ee
Therefore the supergravity D-term potential agrees indeed with
\eqref{eq:DtermC} for very small supersymmetry breaking angle. 
This generalizes the familiar connection between $\Pi$-stability 
and D-term effects to supergravity theories. In order to complete 
the description of the dynamics, we will focus next on superpotential 
interactions. 

\section{Fluxes, Branes and Superpotential Interactions}

In this section we study superpotential interactions of magnetized 
branes in Calabi-Yau 
orientifolds with background fluxes. Brane-flux superpotentials have been 
first discussed in \cite{Lerche:2002yw,Lerche:2002ck}. Our treatment is 
based on the same idea, although our treatment of compact Calabi-Yau 
situations will be closer to \cite{clemens-2002-,Diaconescu:2005jv}.

Let us first consider magnetized branes in the absence of fluxes. The fluxes
will naturally enter the picture at a later stage. 
Our first task is to identify the lowest lying modes which govern the low energy 
physics in the presence of D-branes. 
The massless fields correspond to marginal deformations of the internal bulk-boundary CFT. 
Suppose we have a D-brane wrapping a holomorphic curve $C$ in a Calabi-Yau threefold $X$. The 
marginal deformations of the bulk-boundary CFT are in one-to-one correspondence with deformations 
of the pair $(X,C)$. The infinitesimal deformations of $X$ are classified by $H^1(X,T_X)$. 
Using a standard spectral sequence, one can show that the space $\IH$ of infinitesimal 
deformations of the pair $(X,C)$ fits in an exact sequence the form 
\be\label{eq:infdefA} 
\xymatrix{
& 0  \ar[r]&  H^0(C, N_{C/X}) \ar[r] &  \IH \ar[r] & H^1(X, T_X) \ar[r]^f &  H^1(C, N_{C/X})\\} 
\ee
where $N_{C/X}$ is the normal bundle to $C$ in $X$. The map $f:H^1(X, T_X) \to H^1(C, N_{C/X})$ 
is induced by the natural projection $T_X \to N_{C/X}$. 

From a physical point of view the first term in \eqref{eq:infdefA}, $H^0(C,N_{C/X})$ 
parameterizes marginal boundary operators. The third term $H^1(X,T_X)$
parameterizes marginal deformations 
of the bulk CFT in the absence of boundaries. It is important to note that not
all these marginal operators 
remain marginal in the bulk-boundary CFT. In fact the exact sequence
\eqref{eq:infdefA} shows that only 
those deformations in $H^1(X,T_X)$ which map to zero in $H^1(C, N_{C/X})$ are
marginal deformations of the bulk-boundary theory. 

In our case the Calabi-Yau threefold $X$ is equipped with a holomorphic involution $\sigma$, 
and and the magnetized branes are wrapped on two disjoint curves 
$C,C'=\sigma(C)$ on $X$. Then the infinitesimal deformations 
are captured by the invariant part of \eqref{eq:infdefA} with respect to
$\sigma$
\be\label{eq:infdefB}
\xymatrix{ 
& 0  \ar[r] & H^0(C, N_{C/X}) \ar[r] &  \IH_+ \ar[r] & H^1(X, T_X)_+
\ar[r]^{f_+} &  H^1(C, N_{C/X}).\\} 
\ee

Let us denote by $\CN$ a connected component of the moduli space of data $(X,\sigma, C,C')$. 
Note that there is a natural forgetful map $\rho: \CN\to \CM$, where $\CM$ is a connected 
component of the moduli space of Calabi-Yau threefolds $(X,\sigma)$ with
involution. 
At a generic point in $\CN$, $C,C'$ are $(-1,-1)$ curves on $X$, hence 
\[
H^0(C,N_{C/X})=H^0(C',N_{C'/X})=0.
\] 
The only low energy light modes near such a point in the moduli space 
correspond to deformations of $X$ which preserve 
$(\sigma, C,C')$. 
The map $\rho:\CN\to \CM$ is locally finite-to-one near such a point. 
However, the curves $C,C'$ may have nontrivial normal deformations in $X$ for
special values of the complex structure moduli.  
These normal deformations yield new light fields which have to be taken into account in the 
low energy effective action. This behavior is similar in spirit with the Seiberg-Witten solution of 
$N=2$ gauge theories. Around each point, the low energy theory will have an effective superpotential 
which is a holomorphic function of the lightest fields in the spectrum near that point.  

The local expression of the superpotential on $\CN$ is given by a three-chain 
period of the holomorphic 
three-form on $X$ \cite{Donaldson:1996kp,Witten:1997ep}. More precisely, 
the space $\CN$ can be locally identified near 
each point $(X,\sigma,C,C')$ with an open set $\CU$ in the linear space 
\[ 
H^0(C,N_{C/X}) \oplus \hbox{Ker}(f_+). 
\] 
Let us pick a three chain $\Gamma_0$ interpolating between $C,C'$ on $X$ i.e. 
\[ 
\partial \Gamma = C'-C.
\] 
Then we can extend $\Gamma_0$ to a multivalued family of three-chains $\Gamma_u$, $u\in \CU$ so that 
\[ 
\partial \Gamma_u = C_u - C'_u
\]
for each $u\in \CU$ \cite{clemens-2002-}. 
This extension is obtained by transporting the three-chain 
$\Gamma_0$ to any point in $\CU$ using the Gauss-Manin connection. 
The superpotential is a holomorphic function on $\CU$ given by 
\be\label{eq:locsuperA}
W = \int_{\Gamma_u} \Omega_{X_u} 
\ee
where $\Omega_{X_u}$ is the global holomorphic three-form on $X_u$. Since the overall 
normalization of $\Omega_{X_u}$ is not fixed, \eqref{eq:locsuperA} actually defines a 
local section of the line bundle $\rho^*\CL$ over $\CU$. 

Note that the expression \eqref{eq:locsuperA} is ambiguous since the chain 
$\Gamma_0$ is only defined up to a shift 
\be\label{eq:shift} 
\Gamma_0 \to \Gamma_0 + \gamma.
\ee
where $\gamma$ is a closed three-cycle on $X$. 
This is not a problem from a mathematical point of view since one can show
that the critical set of $W$ is independent of the choice of $\Gamma_0$. 
Nevertheless this ambiguity has a very natural physical interpretation
because we can interpret a shift of the form \eqref{eq:shift} as a 
shift in the background RR flux. 
More precisely, note that the shift \eqref{eq:shift} changes the
superpotential \eqref{eq:locsuperA} by 
\[
\Delta W = \int_{\gamma_u}  \Omega_{X_u}
\]
where $\gamma_u$ is again a family of three-cycles obtained by parallel transport with respect 
to the Gauss-Manin connection. Therefore, using Poincar\'e duality, we can identify 
the ambiguity in the choice of $\Gamma_0$ with a shift 
\[ 
F \to F + \eta
\]
in the background RR flux $F$ on $X$, 
where $\eta\in H^3(X,\IZ)$ is the Poincar\'e dual of $\gamma$.  
This identification is natural since in the presence of D-branes, 
the RR flux is not well defined as 
an element of $H^3(X,\IZ)$; 
an element of $H^3(X,\IZ)$ only determines a shift in the background flux, 
but the overall value of the flux depends on the choice of a trivialization of 
the D-brane charge
\cite{Moore:1999gb}. 
More formally, the RR fluxes take values in a torsor over $H_3(X,\IZ)$.    

We are therefore led to the conclusion that in superstring compactifications,
the superpotential \eqref{eq:locsuperA} should be interpreted as a combined 
brane -- RR flux superpotential. There is no natural way of splitting this 
formula in separate brane and respectively RR flux contributions, but changes 
in the background flux are captured by shifts of the form \eqref{eq:shift}.
Although in this section we have used IIB variables, \eqref{eq:locsuperA}
can be equally interpreted as a IIA superpotential using open string 
mirror symmetry. 


We conclude this discussion with a few remarks. In the next section we
will investigate the vacuum structure of magnetized branes in the octic orientifold 
taking into account both F-term and D-term effects. 

$(i)$ The superpotential \eqref{eq:locsuperA} depends only on the complex
structure deformations of $X$ which preserve the curve $C$. In general these 
deformations span a proper closed subspace of the moduli space $\CM$. We
argued that the 
remaining complex moduli of $X$ are generically massive and do not appear 
in the low energy effective action. This argument is in principle correct at 
generic points in the moduli space, but it may fail at special points 
in the moduli space where the fields we have integrated out become light. 
Such effects can be taken into account extending the superpotential 
\eqref{eq:locsuperA} to a local function of all complex structure moduli. 
Let us consider an open subset $\CV$ of 
\[
H^0(C,N_{C/X}) \oplus H^1(X,T_X)
\]
containing $\CU$ as a closed subset. Then we can use the Gauss-Manin
connection to extend the three-chain $\Gamma_0$ to a family of three-chains 
$\Gamma_v$ labeled by points in $\CV$ and define the extension of $W$ to be 
\be\label{eq:locsuperC} 
W = \int_{\Gamma_v} \Omega_{X_v}. 
\ee
The main difference with respect to the previous case is that the boundary 
of $\Gamma_v$ is no longer a holomorphic cycle on $X_v$ if $v$ is not in
$\CU$. 

$(ii)$ The low energy theory may contain extra light open string fields 
at points in the moduli space $\CN$ where the two curves $C,C'$ coincide. 
Then we will have additional superpotential interactions involving these 
fields as well. 

$(iii)$ The expression \eqref{eq:locsuperA} is very similar to the flux 
superpotential \eqref{eq:IIBsuperA}. In particular they have the same
tree level dependence on the dilaton multiplet $\tau$ and they are 
subject to the same axion shift symmetries. Therefore, using the same low
energy arguments as \cite{Burgess:2005jx} one can show that \eqref{eq:locsuperA} 
is subject to the same nonrenormalization result. This means that this formula 
is reliable at small IIB volume. 

$(iv)$ In general situations, the superpotential \eqref{eq:locsuperA} 
cannot be canonically split into a brane contribution and a flux 
contribution of the form \eqref{eq:IIBsuperA}. However, in special 
cases, this is possible using specific features
of the geometry. For example suppose the threefold $X$ contains a connected 
family of holomorphic curves interpolating between $C,C'$. Then one 
can choose the three-chain $\Gamma$ to be swept by a real one-parameter 
family of holomorphic curves in $X$. It is known that   
the period of $\Omega_X$ on such three-chains vanishes. 
Therefore if we make such a choice, the superpotential \eqref{eq:locsuperA} 
will be identically zero. Then a shift of the form \eqref{eq:shift} will produce 
a superpotential of the form \eqref{eq:IIBsuperA}.

\section{The D-Brane Landscape} 

In this section we explore the magnetized D-brane 
landscape in the octic orientifold model introduced in section 2. 
We compute the F-term and D-term contributions in a neighborhood of the 
Landau-Ginzburg point in the IIA complex structure moduli space $\CK$. 
For technical reasons we will not be able to find explicit solutions 
to the critical point equations. However, given the shape 
of the potential, we will argue that metastable vacuum solutions are 
statistically possible by tuning the values of fluxes. 

Throughout this section we will be working at a generic point in the 
configuration space where all open string fields are massive and can be 
integrated out. Following the reasoning of the previous section, this 
is the expected behavior for D-branes wrapping isolated rigid holomorphic 
curves in a Calabi-Yau threefold. One should however be aware of several 
possible loopholes in this assumption since open string fields may become 
light along special loci in the moduli space.  

In our situation, one should 
be especially careful with the open string-fields in the brane anti-brane 
sector. According to the $\Pi$-stability analysis in section 3, there is a 
tachyonic contribution to the mass of the lightest open string modes
proportional to the phase difference $\theta$. At the same time, we have a 
positive contribution to the mass due to the tension of the string stretching 
between the branes. In order to avoid tachyonic instabilities, we should 
work in a region of the moduli space where the positive contribution is 
dominant. Since the curves are isolated, the positive mass contribution 
is generically of the order of the string scale, which is much larger
than the tachyonic contribution, since $\theta$ is of the order $0.05$. 
Therefore we do not expect tachyonic instabilities in the system as long 
as the moduli are sufficiently generic. 

This argument can be made more precise 
in the mirror IIA picture. As discussed in section 3.2, the IIA description 
of the system involves two disjoint special lagrangian cycles $M,M'$ on 
the Calabi-Yau manifold $Y$. The position of $M,M'$ in $Y$ is determined by 
the calibration conditions \eqref{eq:calib}, which are invariant under a
rescaling of the metric on $Y$ by a constant $\lambda>1$. Such a rescaling 
would also increase the minimal geodesic distance between $Y,Y'$, which 
determines the mass of the open string modes. Therefore, if the volume 
of $Y$ is sufficiently large, we expect the brane anti-brane fields 
to have masses at least of the order of the string scale. 

Even if the open string fields have a positive mass, the system can still be 
destabilized by the brane anti-brane attraction force. Generically, we expect 
this not to be the case as long as the brane-brane fields are sufficiently 
massive since the attraction force is proportional to $\theta$ and it is 
also suppressed by a power of the string coupling. We can understand the
qualitative aspects of the dynamics using a simplified model for the potential 
energy. Suppose that the effective dynamics of the branes can be described 
in terms of a single light chiral superfield $\Phi$. 
Typically this happens when we work near a special point $X_0$ in the complex
structure moduli space where the curves $C,C'$ belong to a one parameter family $\CC$ of
holomorphic curves. The field $\Phi$ corresponds to normal deformations 
of the brane wrapping $C$, which are identified with normal deformations 
of the anti-brane wrapping $C'$ by the orientifold projection. A sufficiently 
generic small complex deformation of $X$ away from $X_0$ induces a mass term
for $\Phi$. Therefore we can model the 
effective dynamics of the system by a potential 
of the form 
\[
m(r-r_0)^2 +c \ln\left({r\over r_0}\right)
\] 
where $r$ parameterizes the separation between the branes. The quadratic 
terms models a mass term for the open string fields corresponding to
normal deformations of the branes in the ambient manifold. The second term 
models a typical two dimensional attractive brane anti-brane potential. 
The constant $c>0$ is proportional to the phase $\theta$ and the 
string coupling $g_s$. 
Now one can check that if $c<< mr_0$, this potential has a local minimum 
near $r=r_0$, and the local shape of the potential near this minimum is 
approximatively quadratic. In our case, we expect $m,r_0$ to be typically 
of the order of the string scale, whereas $c\sim g_s\theta \sim 10^{-2}$ 
therefore the effect of the attractive force is negligible. 

Since it is technically impossible to make these arguments very precise, we
will simply assume that there is a region in configuration space where
destabilizing effects are small and do not change the qualitative behavior 
of the system. Moreover, all open string fields are massive, and we can 
describe the dynamics only in terms of closed string fields. This point 
of view suffices for a statistical interpretation of the D-brane landscape. 
By tuning the values of fluxes, one can in principle explore all regions 
of the configuration space. The vacuum solutions which land outside the 
region of validity of this approximation will be automatically destabilized 
by some of these effects. Therefore there is a natural selection mechanism 
which keeps only vacuum solutions located at a sufficiently generic point in
the moduli space. 

Granting this assumption, we will take the configuration space to be 
isomorphic to the closed string 
moduli space $\CM \times \CK$ described in section 2.1. 
As discussed in section 2.2, we will turn on only RR fluxes $F^A=F_2+F_4+F_6$ 
and NS-NS flux $H^A$. 
In the presence of branes, the NS-NS flux $H^A$ must satisfy the Freed-Witten anomaly 
cancellation condition \cite{Freed:1999vc}, which states that the 
the restriction of $H^A$ to the brane world-volumes $M,M'$ 
must be cohomologically trivial. Taking into account equations \eqref{eq:IIAfluxA},
\eqref{eq:branecharges}, it follows that the integer $q_1$ in
\eqref{eq:IIAfluxA} must be set to zero. Therefore the superpotential 
does not depend on the chiral superfield $\rho$. This can also be seen 
from the analysis of supergravity D-terms in section 3.2. The $U(1)$
gauge group acts as an axionic shift symmetry on $\rho$, therefore 
gauge invariance rules out any $\rho$-dependent terms in the superpotential 
\cite{Villadoro:2005yq}. The connection between the Freed-Witten anomaly condition and 
supergravity has been observed before in \cite{Camara:2005dc}. 

The total effective superpotential is then given by 
\be\label{eq:totsuperB}
W = \int_\Gamma \Omega_X -2p^0\tau,  
\ee 
where $\Gamma$ is a three-chain on $X$ interpolating between the 
two curves $C,C'$. As explained in remark (i) section 4, this expression 
makes sense over the entire moduli space $\CM$ although some complex 
structure deformations may not preserve the curves $C,C'$. This only 
means that some complex 
structure moduli fields are actually massive, and their mass terms 
are encoded in $W$. 
Alternatively, one can take the configuration space to be of the 
form $\CN \times \CK$ by integrating out the massive fields, but 
the two points of view are equivalent, at least generically. 

The F-term contribution to the potential energy is 
\be\label{eq:Fterms} 
U_F = e^K\left(g^{i{\bar\jmath}}(D_iW)(D_{\bar\jmath}{\overline W})+g^{a{\bar b}} 
(D_aW)(D_{\bar b}{\overline W}) -3|W|^2\right).
\ee
where $i,j,\ldots$ label complex coordinates on $\CM$ and  
and $a,b=\rho,\tau$ label complex coordinates on $\CK$. 
The D-term contribution is given by equation \eqref{eq:Denergy}. 
We reproduce it below for convenience 
\[
U_D = {1\over 4p \im(\tau)^3}\left[ 
{\re(Z^0)\im(Z^1) \over \im(Z^1)\re(\CF_1)-\re(Z^0)\im(\CF_0)}\right]^2.
\]
Since the moduli space of the theory is a direct product 
$\CK \times \CM$, the \kah\ potential $K$ in \eqref{eq:Fterms}
is 
\[ 
K = K_\CK + K_\CM.  
\] 
Note that we \kah\ potentials $K_{\CK}$, $K_\CM$ satisfy the following noscale 
relations \cite{Grimm:2004ua}
\be\label{eq:noscale}
g^{i{\bar j}} \partial_i K_{\CM} \partial_{\bar j}K_{\CM} =3 \qquad 
g^{a{\bar b}} \partial_a K_\CK \partial_{\bar b} K_\CK =4.
\ee
Using equations \eqref{eq:holcoordA} and \eqref{eq:kahlerC}, we have 
\[ 
e^{K_{\CK}} = {1\over 4\im(\tau)^4} 
\left[{\re(Z^0)^2\over \im(Z^1)\re(\CF_1)-\re(Z^0)\im(\CF_0)}\right]^2.
\]

Now we have a complete description of the potential energy of the 
system. Finding explicit vacuum solutions using these equations seems 
to be a daunting computational task, given the complexity of the problem. We can however 
gain some qualitative understanding of the resulting landscape by analyzing 
the potential energy in more detail. 

First we have to find a convenient coordinate system on the moduli space
$\CK$. Note that the potential energy is an implicit function of the 
holomorphic coordinates $(\tau,\rho)$ via relations \eqref{eq:holcoordA}. 
One could expand it as a power series in $(\tau,\rho)$, but this would 
be an awkward process. Moreover, the axion ${\widetilde \xi}_1 = \im(\rho)$ 
is eaten by the $U(1)$ gauge field, and 
does not enter the expression for the potential.
Therefore it is more natural to work in coordinates $(\tau,\alpha)$ where 
$\alpha$ is the algebraic coordinate on the underlying $N=2$ \kah\ moduli 
space. As explained in section 2.1, $\alpha$ takes real values in the 
orientifold theory. 

There is a more conceptual reason in favor of the coordinate $\alpha$ instead 
of $\rho$, namely $\alpha$ is a coordinate on the Teichm\"uller space of $Y$ rather 
than the complex structure moduli space. Since in the $\Pi$-stability
framework the phase of the central charge is defined on the Teichm\"uller
space, $\alpha$ is the natural coordinate when D-branes are present. 

Next, we expand the potential energy in terms of $(\tau,\alpha)$ using the 
relations \eqref{eq:holcoordA}. Dividing the two equations in
\eqref{eq:holcoordA}, we obtain
\be\label{eq:ratioA}
{\rho + {\overline \rho}\over \tau - {\overline \tau}} = 
2i {\re(\CF_1)\over \re(Z^0)} 
\ee
Let us denote the ratio of periods in the right hand side of equation
\eqref{eq:holcoordA} by 
\be\label{eq:ratioB}
R(\alpha) = {\re(\CF_1)\over \re(Z^0)}.
\ee
Using equations \eqref{eq:ratioA} and \eqref{eq:ratioB}, 
we find the following relations 
\be\label{eq:derivativesA} 
{\partial \alpha \over \partial \rho } = {1\over 2i} {1\over \tau -{\overline
\tau}}\left({\partial R\over \partial \alpha} \right)^{-1} \qquad 
{\partial \alpha \over \partial \tau } = - {R\over \tau -{\overline
\tau}}\left({\partial R\over \partial \alpha} \right)^{-1}.
\ee
Now, using the chain differentiation rule, we can compute the derivatives 
of the \kah\ potential as functions of $(\tau, \alpha)$. Let us introduce the
notation 
\[ 
V(\alpha) = {\im(Z^1)\re(\CF_1)-\re(Z^0)\im(\CF_0)\over \re(Z^0)^2}.
\]
Then we have 
\be\label{eq:kahderA}
\bal 
\partial_\tau K_\CK & =-\partial_{\bar \tau}K_\CK= -{2\over \tau -{\overline \tau}}\left[2 -R
  {\partial_\alpha V\over V} \left({\partial_\alpha  R}
  \right)^{-1}\right]\cr
\partial_\rho K_\CK & = \partial_{\bar \rho} K_\CK= {i\over \tau-{\overline \tau}}
{\partial_\alpha V\over V} \left({\partial_\alpha R}
  \right)^{-1}\cr
\partial_{\tau {\bar \tau}} K_\CK & = -{2\over (\tau -{\overline \tau})^2}
\left[2 -R{\partial_\alpha V\over V} \left({\partial_\alpha  R}
  \right)^{-1}-R\partial_\alpha \left(R{\partial_\alpha V\over
      V}(\partial_\alpha R)^{-1}\right)(\partial_\alpha R)^{-1}\right]\cr
\partial_{\tau{\bar \rho}}K_\CK& = -\partial_{\rho{\bar \tau}}K_\CK = 
-{i\over (\tau - {\overline \tau})^2} \left[{\partial_\alpha V\over V} \left({\partial_\alpha  R}
  \right)^{-1}+R\partial_\alpha \left({\partial_\alpha V\over
      V}(\partial_\alpha R)^{-1}\right)(\partial_\alpha R)^{-1}\right]\cr
\partial_{\rho{\bar \rho}}K_\CK & = {1\over 2(\tau - {\overline \tau})^2} 
\partial_\alpha \left({\partial_\alpha V\over
      V}(\partial_\alpha R)^{-1}\right)(\partial_\alpha R)^{-1}\cr
\eal
\ee
Using equations \eqref{eq:kahderA}, and the power series expansions of the
periods computed in appendix A, we can now compute the expansion of the
potential energy as in terms of $(\tau,\alpha)$. The D-term contribution 
takes the form 
\be\label{eq:DtermH} 
U_D = {1\over p \im (\tau)^3}(0.03125-0.00178\alpha^2 +0.00005\alpha^4 + \ldots ).
\ee
We will split the F-term contribution into two parts
\[ 
U_F = U_F^{\CM}+ U_F^{\CK} 
\]
where 
\[
\bal 
U_F^\CM &= e^{K_\CK+K_\CM}\left(g^{i{\bar\jmath}}(D_iW)(D_{\bar\jmath}{\overline W})-3|W|^2\right)\cr
U_F^\CK &= e^{K_\CK+K_\CM} \left(g^{a{\bar b}}(D_aW)(D_{\bar b}{\overline W})\right).\cr
\eal
\] 
We will also write the superpotential \eqref{eq:totsuperB} in the form 
\[ 
W = W_0(z^i) +k \tau 
\] 
where $k=-2p^0$. 
The factor $e^{K_\CK}$ and the inverse metric coefficients $g^{a{\bar b}}$ can
be expanded in powers of $\alpha$ using the equations \eqref{eq:kahderA} and 
formulas \eqref{eq:perexp} in appendix A. Using the noscale relations \eqref{eq:noscale}, 
we find the following expressions 
\be\label{eq:FtermB} 
\bal 
& U_F^{\CM} =  {e^{K_\CM}\over 4\im(\tau)^4}
(0.0625 -0.00357\alpha^2 + 0.00004 \alpha^4 + \ldots ) \cr
&  
\left(g^{i{\bar\jmath}}(\partial_i W_0) (\partial_{\bar\jmath} {\overline W}_0)+ 
g^{i{\bar\jmath}}[(\partial_i W_0)(\partial_{\bar\jmath} K_\CM) ({\overline W}_0 + k
  {\overline \tau})+ (\partial_{\bar\jmath}{\overline W}_0)( \partial_i K_\CM)
  (W_0+k\tau)] \right)
\eal
\ee
\be\label{eq:FtermC} 
\bal 
U_F^{\CK} = &{e^{K_\CM}\over \im(\tau)^4}
\bigg[\im(\tau)^2(0.03125-0.00073\alpha^2 +0.00001\alpha^4+\ldots)k^2 \cr
& -\im(\tau)(0.03125 -0.00178\alpha^2 +0.00002\alpha^4+\ldots)(2k^2\im(\tau) +
2k\im(W_0))\cr
& +(0.0625 -0.00357\alpha^2 +0.00004\alpha^4+\ldots)(k^2\tau{\overline \tau}
+k\tau{\overline W_0} + k{\overline \tau}W_0 + |W_0|^2)\bigg]\cr
\eal
\ee
Let us now try to analyze the shape of the landscape determined by the
equations \eqref{eq:DtermH} and  \eqref{eq:FtermB}, \eqref{eq:FtermC}. 
We rewrite the contribution \eqref{eq:FtermB} to the potential 
energy in the form 
\be\label{eq:FtermD} 
\bal 
& U_F^{\CM} =  {e^{K_\CM}\over 4\im(\tau)^4}
(0.0625 -0.00357\alpha^2 + 0.00004 \alpha^4 + \ldots ) 
\left(P + kM\re(\tau) + kN \im(\tau)\right)\cr
\eal
\ee
where 
\be\label{eq:FtermE} 
\bal 
& P = 
g^{i{\bar\jmath}}(\partial_i W_0) (\partial_{\bar\jmath} {\overline W}_0)+ 
g^{i{\bar\jmath}}[(\partial_i W_0)(\partial_{\bar\jmath} K_\CM) {\overline W}_0 + 
(\partial_{\bar\jmath}{\overline W}_0)( \partial_i K_\CM)W_0] \cr
& M = g^{i{\bar\jmath}}[(\partial_i W_0)(\partial_{\bar\jmath} K_\CM) + 
(\partial_{\bar\jmath}{\overline W}_0)( \partial_i K_\CM)]\cr
& N = (-i)g^{i{\bar\jmath}}[(\partial_i W_0)(\partial_{\bar\jmath} K_\CM) -
(\partial_{\bar\jmath}{\overline W}_0)( \partial_i K_\CM)]\cr
\eal
\ee
Then the $\alpha$ expansion of the F-term  potential energy can be written 
as 
\[ 
U_F = U_F^{(0)} + \alpha^2 U_F^{(2)} + \ldots 
\]
where 
\be\label{eq:FtermF} 
\bal 
U_F^{(0)} = 0.0156{e^{K_\CM}\over \im(\tau)^4}
[& P+k(N+4\im(W_0))\im(\tau)+2k^2\im(\tau)^2\cr 
& + 4|W_0|^2 + k(M+8\re(W_0))\re(\tau)+4k^2 \re(\tau)^2]\cr 
\eal
\ee
\be\label{eq:FtermG} 
\bal 
U_F^{(2)} = -0.00178{e^{K_\CM}\over 2 \im(\tau)^4}
[& P + k(N+4\im(W_0))\im(\tau) + 0.82 k^2\im(\tau)^2 \cr
& + 4|W_0|^2 + k(M+8\re(W_0))\re(\tau)+4k^2 \re(\tau)^2]\cr 
\eal 
\ee

The critical point equations resulting from \eqref{eq:DtermH},
\eqref{eq:FtermB} and \eqref{eq:FtermC} are very complicated, and 
we will not attempt to find explicit solutions. We will try to gain 
some qualitative understanding of the possible solutions exploiting 
some peculiar aspects of the potential. Note that all contributions to the 
potential energy depend on even powers of $\alpha$. Then it is obvious 
that $\alpha=0$ is a solution to the equation 
\[ 
\partial_\alpha U =0 
\] 
where $U = U_D + U_F^\CM + U_F^\CK$. Moreover we also have 
\[ 
(\partial_i\partial_\alpha U)_{\alpha=0} = (\partial_\tau \partial_\alpha
U)_{\alpha=0} =0.
\] 
This motivates us to look for critical points with $\alpha=0$. Then, the
remaining critical point equations are 
\be\label{eq:crtptA}
(\partial_i U)_{\alpha=0} = (\partial_\tau U)_{\alpha=0} =0 
\ee
plus their complex conjugates.  

The second order coefficient of $\alpha$ in the total potential energy 
is 
\be\label{eq:alphamass} 
U_F^{(2)} -0.00178{1\over p \im(\tau)^3 }. 
\ee
Since the mixed partial derivatives are zero at $\alpha=0$,  
in order to obtain a local minimum, the expression \eqref{eq:alphamass} 
must be positive. This is a first constraint on the allowed solutions to 
\eqref{eq:crtptA}.  

Next, let us examine the $\tau$ dependence of the potential for $\alpha=0$
and fixed values of the \kah\ parameters. Note that $U_F^{(0)}$ given by 
equation \eqref{eq:FtermF} is a 
quadratic function of the axion $\re(\tau)$. 
For any fixed values of $\im(\tau)$ and the \kah\ parameters, this 
function has a minimum at 
\be\label{eq:realtau}
\re(\tau) = -{8\re(W_0) +M\over 8k}. 
\ee 
Therefore we can set $\re(\tau)$ 
to its minimum value in the potential energy, obtaining an effective potential 
for the \kah\ parameters and the dilaton $\im(\tau)$. Then equations
\eqref{eq:FtermF}, \eqref{eq:FtermG} become 
\be\label{eq:FtermH} 
\bal 
U_F^{(0)} = 0.0156{e^{K_\CM}\over \im(\tau)^4}
\bigg[& P+k(N+4\im(W_0))\im(\tau)+2k^2\im(\tau)^2\cr 
& + 4|W_0|^2 - {1\over 16} (M+8\re(W_0))^2\bigg]\cr 
\eal
\ee
\be\label{eq:FtermI} 
\bal 
U_F^{(2)} = -0.00178{e^{K_\CM}\over 2 \im(\tau)^4}
\bigg [& P + k(N+4\im(W_0))\im(\tau) + 0.82 k^2\im(\tau)^2 \cr
& + 4|W_0|^2 - {1\over 16} (M+8\re(W_0))^2 \bigg]\cr 
\eal 
\ee
Now let us analyze the dependence of $U_F^{(0)}$ on $\im(\tau)$. It will be
more convenient to make the change of variables 
\[
u = {1\over \im(\tau)}
\] 
since $u$ is proportional to the string coupling constant. Then 
$U_F^{(0)}$ becomes a quartic function of the form 
\be\label{eq:FtermJ} 
U_F^{(0)} = Au^2 -Bu^3 +Cu^4 
\ee
where 
\be\label{eq:quartcoeff} 
\bal 
& A = 2k^2 \cr
& B = -kN - 4k\im(W_0)\cr
& C = P + 4|W_0|^2 - {1\over 16}(M+8\re(W_0))^2. \cr
\eal 
\ee
The behavior of this function for fixed values of the \kah\ parameters 
is very simple. For positive $A$, this function has a local minimum away 
from the origin if and only if the following inequalities are satisfied 
\be\label{eq:dilatonA} 
B>0 \qquad C>0 \qquad \hbox{and} \qquad 9B^2 > 32AC.
\ee 
The minimum is located at 
\be\label{eq:dilatonB} 
u_0 = {3B+\sqrt{9B^2-32AC}\over 8C}. 
\ee
Therefore, in order to construct metastable vacua, we have to find solutions
to the equations \eqref{eq:crtptA} satisfying the inequalities
\eqref{eq:dilatonA}. Moreover, we would like $u_0$ to be small in order 
to obtain a weakly coupled theory. The conditions \eqref{eq:dilatonA} 
translate to  
\be\label{eq:dilatonC}
\bal 
9(N+4\im(W_0))^2 & > 64
\left(P+4\im(W_0)^2 - M\re(W_0) - {M^2\over 16}\right) >0\cr
& k\left(N+\im(W_0)\right)  < 0 \cr
\eal 
\ee
where $P,M,N$ are given by \eqref{eq:FtermE}. 
This shows that we need a certain amount of fine tuning of the background 
RR fluxes in order to obtain a metastable vacuum. Note that in our construction 
the fluxes are not constrained by tadpole cancellation conditions, therefore 
we can tune them at will. Statistically, this improves our chances of finding 
a solution with the required properties. 

Finally, note that we have to impose one more condition, namely the 
second order coefficient \eqref{eq:alphamass} in the $\alpha$ expansion 
of the potential should be positive. Assuming that we have found a solution 
of equations \eqref{eq:crtptA} which stabilizes $u$ at the value
$0<u_0<1$, let us compute this coefficient as a function of $(u_0,A,B,C)$.  
Note that equation \eqref{eq:FtermI} can be rewritten as 
\be\label{eq:FtermK} 
U_F^{(2)} = 0.00178{e^{K_\CM}\over 2}(Bu_0^3 -0.4 Au_0^2 -Cu_0^4).
\ee 
Equation \eqref{eq:dilatonB} yields 
\be\label{eq:subst}
B = {4\over 3} Cu_0 +{2\over 3} {A\over u_0}
\ee
Substituting \eqref{eq:subst} in \eqref{eq:FtermK}, and adding the D-term 
contribution, the coefficient of $\alpha^2$ becomes 
\be\label{eq:alphamassB} 
0.00178\left[{e^{K_\CM}}({2\over 15}Au_0^2 + {1\over 6}Cu_0^4) - {1\over p}
  u_0^3\right] 
\ee
Since $C>0$, a sufficient condition for \eqref{eq:alphamassB} to be positive is 
\be\label{eq:alphamassC} 
{2p\over 15} Ae^{K_\CM} > u_0 \quad \Rightarrow \quad {4pk^2\over 15} > u_0  
\hbox{vol}(Y).
\ee
Here we have used 
\[ 
e^{K_\CM} = {1\over \hbox{vol}(Y)}.
\]
This condition reflects the fact that the F-term and D-term contributions to
the potential energy must be of the same order of magnitude in order 
to obtain a metastable vacuum solution. If the volume of $Y$ is too 
large, there is a clear hierarchy of scales between the two contributions, 
and the D-term is dominant. This would give rise to a runaway behavior 
along the direction of $\alpha$. On the other hand, we have to make sure 
that the volume of $Y$ is sufficiently large so that the IIA supergravity 
approximation is valid. Therefore some additional amount of fine 
tuning is required in order to obtain a reliable solution. 

In conclusion, metastable nonsupersymmetric vacua at $\alpha=0$ can be in
principle obtained by tuning the IIA RR flux $F^{(A)}$ and NS-NS flux
$H^A=k\beta^0$ so that conditions \eqref{eq:dilatonC}, \eqref{eq:alphamassC} 
are satisfied at the critical point. A more precise statement would require 
a detailed numerical analysis, which we leave for future work. 

We would like to conclude this section with a few remarks. 

$(i)$ In this paper we have taken a conservative approach towards fluxes,
avoiding half flat structures in the IIA theory, which correspond to IIB 
NS-NS flux $H^B$. If one is willing to consider 
compactifications of this form, we have additional terms in the
superpotential. In IIB variables, these terms would read 
\[ 
-\tau \int_X \Omega_X \wedge H^B.
\]
One can also turn on additional flux degrees of freedom as advocated in 
\cite{Shelton:2005cf,Aldazabal:2006up}.
Such terms may be helpful in the above fine tuning process.

$(ii)$ We have also restricted ourselves to singly wrapped magnetized 
D5-branes. One could in principle consider multiply wrapped D5-branes 
as long we can maintain the phase difference $\theta$ sufficiently small. 
If this is possible, we would obtain an additional nonperturbative 
contribution to the superpotential of the form 
\[ 
b e^{-a\tau}. 
\]
Such terms may be also helpful in the fine tuning process. 

$(iii)$ Finally, note that we could also allow for a nonzero 
background value of the RR zero-form $F_0$, which was also 
set to zero in this paper. Then, according to \cite{DeWolfe:2005uu}, 
there is an additional contribution to the RR tadpole cancellation 
condition, which becomes 
\[ 
p -km_0 -|N_{O3}|=0. 
\] 
If we choose $k,m_0$ so that $km_0>0$, it follows that $p$ can be 
larger than $|N_{O3}|$. In fact it seems that there is no upper 
bound on $p$, hence we could make the supersymmetry 
breaking D-term very small by choosing a large $p$. 
This may have important consequences for the scale of 
supersymmetry breaking in string theory. 

$(iv)$ Note that the vacuum construction mechanism proposed above 
can give rise to de Sitter or anti de Sitter vacua, depending on the 
values of fluxes. In particular, it is not subject to the no-go 
theorem of \cite{Maldacena:2000mw} because the magnetized branes 
give a positive contribution to the potential energy. 
In principle we could try to employ the same strategy 
in order to construct nonsupersymmetric metastable vacua of the
F-term potential energy \eqref{eq:Fterms} in the 
absence of magnetized branes. Then we have several options for 
RR tadpole cancellation. We can turn on background $F_0$ 
flux as in \cite{DeWolfe:2005uu} or local tadpole cancellation 
by adding background D6-branes. It would be interesting to 
explore these alternative constructions in more detail. 

$(v)$ Since it is quite difficult to find explicit vacuum solutions, 
it would be very interesting to attempt a systematic statistical 
analysis of the distribution of vacua along the lines of 
\cite{Ashok:2003gk,douglas-2004-,Denef:2004cf,Denef:2004ze,Douglas:2004zg,
DeWolfe:2004ns,Acharya:2005ez}.

$(vi)$ In our approach the scale of supersymmetry breaking is essentially 
determined by the total RR tadpole $p=|N_{O3}|$ of the orientifold model. 
While this tadpole is typically of the order of $32$ in perturbative models, 
it can reach much higher values in orientifold limits of F-theory. It would 
be very interesting to implement our mechanism in such an F-theory 
compactification, perhaps in conjunction with other moduli 
stabilization mechanism
\cite{Acharya:2002kv,Kachru:2003aw,Denef:2004dm,Denef:2005mm,Lust:2005dy,
Reffert:2005mn,Balasubramanian:2005zx}.
Provided that the dynamics can still be kept under control, we would 
then obtain smaller supersymmetry breaking scales. 

\appendix 
\bigskip 
\section{$\Pi$-Stability on the Octic and $N=2$ \kah\ Moduli Space} 

In this appendix we analyze the $N=2$ \kah\ moduli space and stability 
of magnetized branes for the octic hypersurface. Recall \cite{Klemm:1992tx} 
that the mirror family is described by the equation 
\be\label{eq:mirroctA}
x_1^8+x_2^8 + x_3^8 + x_4^8 +x_5^2 -\alpha x_1x_2x_3x_4x_5 =0. 
\ee
in $WP^{1,1,1,1,4}/(\IZ_8^2 \times \IZ_2)$. 
The moduli space of the mirror family can be identified with a 
sector in the $\alpha$ plane defined by 
\[ 
0\leq \hbox{arg}(\alpha) < {2\pi \over 8}.
\]
The entire $\alpha$ plane contains eight such sectors, which are 
permuted by monodromy transformations about the LG point $\alpha=0$. 
In this parameterization, the LCS point is at $\alpha=\infty$, 
and the conifold point is at $\alpha =4$. 

A basis of periods for this family has been computed in \cite{Klemm:1992tx}
by solving the Picard-Fuchs equations. For our purposes it is convenient 
to write the solutions to the Picard-Fuchs equations in integral 
form 
\be\label{eq:octperA}
\bal 
& \Pi_0 = {1\over 2\pi i} \int ds {\Gamma(1+8s)\Gamma(-s)\over 
\Gamma(1+s)^3 \Gamma(1+4s)}e^{i\pi s} (\alpha)^{-8s} \cr
& \Pi_1 = -{1\over (2\pi i)^2 } \int ds {\Gamma(1+8s)\Gamma(-s)^2 \over 
\Gamma(1+s)^2 \Gamma(1+4s)} (\alpha)^{-8s} \cr
& \Pi_2= {2\over (2\pi i)^3} \int ds {\Gamma(1+8s)\Gamma(-s)^3 \over 
\Gamma(1+s)\Gamma(1+4s) } e^{i\pi s} (\alpha)^{-8s} \cr
& \Pi_3 =- {1\over (2\pi i)^4} \int ds {\Gamma(1+8s)\Gamma(-s)^4 \over 
\Gamma(1+4s)} (\alpha)^{-8s}.
\eal 
\ee
as in \cite{Aspinwall:2001dz}.
All integrals in  \eqref{eq:octperA} are contour integrals in the 
complex $s$-plane. The contour runs from $s=-\epsilon -i\infty$ to 
$-\epsilon + i \infty$ along the imaginary axis and it can be closed either 
to the left or to the right. If we close the contour to the right, we obtain 
a basis of solutions near the LCS limit $\alpha=\infty$, while if we close 
the contour to the left, we obtain a basis of solutions near the LG point 
$\alpha=0$. Near the large radius limit it is more convenient to write 
the solutions in terms of the coordinate $z=\alpha^{-8}$. 

Note that there is a different set of solutions at the LG point \cite{Klemm:1992tx}
of the form 
\be\label{eq:octperB}
w_k(\alpha) = \Pi_0(e^{2\pi k i}\alpha), \qquad k =0,\ldots , 7.
\ee
In particular we have an alternative basis $[w_2\ w_1\ w_0\ w_7]^{tr}$ near 
$\alpha=0$. The transition matrix between the two bases is 
\be\label{eq:transmatrix} 
\left[\begin{array}{c} \Pi_0 \cr \Pi_1 \cr \Pi_2 \cr \Pi_3 \end{array}
\right] = 
\left[\begin{array}{cccc} 
0 & 0 & 1 & 0 \cr 
{1\over 2} & {1\over 2} & -{1\over 2} & -{1\over 2}\cr
0 & -1 & -2 & -1 \cr
-1 & -{1\over 2} & {1\over 2} & 1 \cr
\end{array}\right]
\left[\begin{array}{c}
w_2\cr w_1 \cr w_0 \cr w_7 
\end{array} \right]
\ee
In section 2 we have used a third basis of periods $[Z^0\ Z^1\ \CF_1\
\CF_0]^{tr}$ compatible with the orientifold projection. The relation between
the orientifold basis and the LG basis $[w_2\ w_1\ w_0\ w_7]^{tr}$ is given in 
equation \eqref{eq:transmapA}. The power series expansion of the orientifold 
periods at the LG point is 
\be\label{eq:perexp} 
\bal 
\re(Z^0) &=-0.37941 \alpha + 0.00541 \alpha^3 + 0.00009 \alpha^5 + \ldots \cr 
\im(Z^1) &=-0.53656 \alpha + 0.00766 \alpha^3 -0.00012 \alpha^5 + \ldots \cr
\re(\CF_1) & = 1.29538 \alpha -0.00317 \alpha^3 -0.00005 \alpha^5 + \ldots \cr
\im(\CF_0)&=0.31431 \alpha -0.02615 \alpha^3 +0.00043 \alpha^5 + \ldots \cr
\eal 
\ee

Now let us discuss some geometric aspects of octic hypersurfaces required 
for the $\Pi$-stability analysis. For intersection theory computations, 
it will be more convenient to represent $X$ as a hypersurface in a smooth 
toric variety $Z$ obtained by blowing-up the singular point of the weighted 
projective space $WP^{1,1,1,1,4}$. $Z$ is defined by
the following $\IC^\times \times \IC^\times$ action 
\be\label{eq:toricdata} 
\begin{array}{cccccc}
 x_1 & x_2 & x_3 & x_4 & u & v \cr
 1 & 1 & 1 & 1 & -4 & 0 \cr
 0 & 0 & 0 & 0 & 1 & 1 \cr
\end{array}
\ee 
with forbidden locus $\{x_1=x_2=x_3=x_4=0\}\cup \{u=v=0\}$.
The Picard group of $Z$ is generated by two divisor classes $\eta_1,\eta_2$
determined by the equations 
\be\label{eq:picard} 
\eta_1: \ x_1 =0 \qquad \eta_2: \ v =0. 
\ee
The cohomology ring of $Z$ is determined by the relations 
\be\label{eq:relations}
\eta_2^4 =64 \qquad \eta_2(\eta_2-4\eta_1) =0.
\ee
The total Chern class of $Z$ is given by the formula 
\be\label{eq:totchern} 
c(Z) = (1+\eta_1)^4 (1-4\eta_1+\eta_2)(1+\eta_2) 
\ee
and the hypersurface $X$ belongs to the linear system $|2\eta_2|$. 
Using the adjunction formula 
\be\label{eq:adj} 
c(X) = {c(Z)\over (1+2\eta_2)}
\ee
one can easily compute 
\be\label{eq:cherncls}
c_1(X) =0 \qquad c_2(X) = 22\eta_1^2 \qquad \td(X) = 1 +{11\over 6}\eta_1^2. 
\ee
Note that the divisor class $\eta_2-4\eta_1$ has trivial restriction to 
$X$, therefore the Picard group of $X$ has rank one, as expected. 
A natural generator is $\eta_1$, which can be identified with a hyperplane 
section of $X$ in the weighted projective space $WP^{1,1,1,1,4}$. 
Then we will write the complexified \kah\ class as $B+iJ = t
\eta_1$. For future reference, note that we will denote by $E(p)$ the 
tensor product $E\otimes \CO_X(p\eta_1)$ for any sheaf (or derived object) 
$E$ on $X$. 

Employing the conventions of \cite{Aspinwall:2002ke}, we will define the central 
charge of a D-brane $E$ in the large radius limit to be 
\be\label{eq:braneZ} 
Z^\infty(E) = \int_X e^{B+iJ} \ch(E) \sqrt{\td(X)}. 
\ee
This is a cubic polynomial in $t$. Using the mirror map
\be\label{eq:mirrormap}
t = {\Pi_1\over \Pi_0}
\ee
and the asymptotic form of the periods 
\be\label{eq:asymp} 
\bal 
\Pi_1 & = t + \ldots \cr
\Pi_2 & = t^2 +t -{11\over 6} +\ldots \cr
\Pi_3 & = {1\over 6}{t^3} -{13\over 12}t + \ldots \cr
\eal
\ee
we can determine the exact expression of the period $Z_E$ as a function 
of the algebraic coordinate $\alpha$. 
The phase of the central charge is defined as
\be\label{eq:phasedef} 
\phi(E) = -{1\over \pi}\hbox{arg}(Z(E))
\ee
and is normalized so that it takes values $ -2 < \phi(E) \leq 0$ at 
the large radius limit point. 

As objects in the derived category $D^b(X)$, the magnetized branes are 
given by 
\be\label{eq:derbranes}
{\underline {\CO_C}}(p-1)\qquad {\underline {\CO_{C'}}}(-p-1)[1]
\ee
where $C,C'$ are smooth rational curves on $X$ conjugated under the holomorphic involution.
Given a coherent sheaf $E$ on $X$, we have denoted by ${\underline {E}}$ 
the one term complex which contains $E$ in degree zero, all other terms being 
trivial. In order to compute their asymptotic central charges using 
formula \eqref{eq:braneZ}, we have to use the Grothendieck-Riemann-Roch 
theorem for the embeddings $\iota:C \to X$, $\iota':C'\to X$. Since the
computations are very similar, it suffices to present the details only for 
one of these objects, for example the first brane in \eqref{eq:derbranes}. 

Given a line bundle $\CL\to C$, the Chern character of its pushforward  
$\iota_*(\CL)$ to $X$ is given by 
\be\label{eq:GRRA} 
\ch(\iota_*(\CL))\td(X) = \iota_*(\ch(\CL)\td(C)).
\ee
In our case \eqref{eq:GRRA} yields 
\be\label{eq:GRRB} 
\ch_0(\iota_*(\CL))=\ch_1(\iota_*(\CL))=0\qquad 
\ch_2(\iota_*(\CL))= [C]\qquad \ch_3(\iota_*(\CL))= (\hbox{deg}(\CL)+1)[pt]
\ee 
where $[C]\in H^{2,2}(X)$ denotes the Poincar\'e dual of $C$ and $[pt]\in H^{3,3}(X)$
denotes the Poincar\'e dual of a point on $X$. 
The shift by $1$ in $\ch_3(\iota_*(\CL))$ represents the contribution of the 
Todd class of $C$ 
\[
\hbox{\td}(C)= 1 + \half c_1(C) 
\]
to the right hand side of equation \eqref{eq:GRRA}. 
From a physical point of view, this can be thought of as D3-brane charge 
induced by a curvature effect. 
Using formulas \eqref{eq:braneZ}, \eqref{eq:GRRB} it is easy to compute 
\be\label{eq:asympbraneZ} 
Z^\infty\left({\underline {\CO_C}}(p-1)\right) = t + p \qquad 
Z^\infty\left({\underline {\CO_{C'}}}(-p-1)[1]\right)= -t +p.
\ee
The exact expressions for the central charges are 
\be\label{eq:exactbraneZ}
Z\left({\underline {\CO_C}}(p-1)\right) = \Pi_1 + p\Pi_0 \qquad 
Z\left({\underline {\CO_{C'}}}(-p-1)[1]\right)= -\Pi_1 +p\Pi_0.
\ee
Taking into account the transition matrices \eqref{eq:transmapA}, 
\eqref{eq:transmatrix}, it is clear that these formulas are identical 
with \eqref{eq:centralchB} in the main text. 
In order to study the behavior of their phases near the LG point, 
we have to rewrite the central charges \eqref{eq:exactbraneZ} 
in terms of the basis $[w_2\ w_1\ w_0\ w_7]^{tr}$ using the transition 
matrix \eqref{eq:transmatrix}. We find 
\be\label{eq:LGZA} 
\bal 
& Z\left({\underline {\CO_C}}(p-1)\right) = \half(w_2+w_1-w_0-w_7)+ pw_0 \qquad 
\cr & Z\left({\underline {\CO_{C'}}}(-p-1)[1]\right)= -\half(w_2+w_1-w_0-w_7)+pw_0.
\cr \eal \ee 
Note that the central charge of a single D3-brane is 
\be\label{eq:threebrane} 
Z({\underline {\CO_{pt}}}) = w_0.
\ee
Then, using the expansions \eqref{eq:perexp} we can plot the relative phase 
\be\label{eq:relphaseD} 
\theta = \phi\left({\underline {\CO_C}}(p-1)\right)-\phi({\underline
  {\CO_{pt}}}) 
\ee
near the LG point, obtaining the graph in figure 1. 

In the remaining part of this section, we will address the question of 
stability of magnetized brane configurations near the LG point. As explained 
below figure 1, we will analyze stability with respect to the most natural 
decay channels from the geometric point of view.
We will show below that the objects \eqref{eq:derbranes} 
are stable with respect to all such decay processes, which is strong 
evidence for their stability at the LG point. 
Since all these computations are very similar, it suffices to  consider only one 
case in detail. For the other cases we will just give the final results. 
 
Decay channels in the $\Pi$-stability framework are classified by triangles
in the derived category \cite{Aspinwall:2001dz}. In our case, the most natural decay
channels are in fact determined by short exact sequences of sheaves. 
For example let us consider the following short exact sequence 
\be\label{eq:decayA} 
0 \to \CJ_C(p-1) \to \CO_X(p-1)\to \CO_{C}(p-1) \to 0 
\ee
where $\CJ_C$ is the ideal sheaf of $C$ on $X$.  
The first two terms represent rank one D6-branes on $X$ with lower 
D4 and D2 charges. All three terms are stable BPS states in the 
large volume limit. The mass of the lightest open string states stretching 
between the first two branes in the sequence \eqref{eq:decayA} is determined 
by the relative phase
\be\label{eq:tachyonmassA}
\Delta \phi = \phi\left({\underline {\CO_X}}(p-1)\right)
-\phi\left({\underline {\CJ_C}}(p-1)\right). 
\ee 
If $\Delta\phi <1$, the lightest state in this open string sector 
is tachyonic, and these two branes will form a bound state isomorphic 
to ${\underline {\CO_C}}(p-1)$ by tachyon condensation. In this case 
${\underline {\CO_C}}(p-1)$ is 
stable. If $\Delta \phi>1$, the lightest open string state has positive 
mass, and it is energetically favorable for ${\underline {\CO_{C}}}(p-1)$ 
to decay into 
${\underline {\CJ_C}}(p-1)$ and ${\underline {\CO}}_X(p-1)$. 
In this case ${\underline {\CO_C}}(p-1)$ is unstable. 
Therefore we have to compute the phase difference $\Delta \phi$ as a 
function of $\alpha$ in order to find out if this decay takes place 
anywhere on the real $\alpha$ axis. For the purpose of this computation 
it is more convenient to denote $q=p-1$, and perform the calculations 
in terms of $q$ rather than $p$. 

We have 
\be\label{eq:chernchar}
\bal
Z^\infty\left({\underline {\CO}}_X(q)\right)& =\int_X e^{(t+q)\eta_1} \sqrt{\hbox{Td}(X)} \cr
& = {1\over 3}(t+q)^3 +{11\over 6}(t+q) \cr
Z^\infty\left({\underline {\CJ_C}}(q)\right)& = 
\int_X e^{(t+q)\eta_1} \sqrt{\hbox{Td}(X)} - Z^\infty\left({\underline {\CO_C}}(q)\right)\cr
& = {1\over 3}(t+q)^3 +{5\over 6}(t+q)-1.\cr
\eal
\ee 
Using the asymptotic form of the periods \eqref{eq:asymp} and 
formulas \eqref{eq:chernchar}, we find the following expressions 
for the exact central charges 
\be\label{eq:exactZA} 
\bal 
Z\left({\underline {\CO}}_X(q)\right)& =2\Pi_3 +q\Pi_2 +(q^2-q+4)\Pi_1 +
\left({1\over 3}{q^3} +{11\over 3}q\right)\Pi_0 \cr
Z\left({\underline {\CJ_C}}(q)\right)& = 2\Pi_3 +q\Pi_2 +(q^2-q+3)\Pi_1 +
\left({1\over 3}{q^3} +{8\over 3}q-1\right)\Pi_0\cr
\eal
\ee 
In terms of the LG basis of periods, these expressions read 
\be\label{eq:exactZB}
\bal 
Z\left({\underline {\CO}}_X(q)\right)& =\left(\half q^2-\half q\right)w_2
+\left(\half q^2 -{3\over 2} q +1\right)w_1 \cr
& \qquad + \left({1\over 3}q^3 -\half q^2
  +{13\over 6}q -1 \right)w_0 +\left(-\half q^2 -\half q\right) w_7 \cr
Z\left({\underline {\CJ_C}}(q)\right)& = \left(\half q^2-\half q-\half \right)w_2
+\left(\half q^2 -{3\over 2} q +\half \right)w_1 \cr
& \qquad + \left({1\over 3}q^3 -\half q^2
  +{7\over 6}q -{3\over 2} \right)w_0 +\left(-\half q^2 -\half q +\half \right) w_7\cr
\eal
\ee 
Substituting the expressions \eqref{eq:octperA} in \eqref{eq:exactZA},
\eqref{eq:exactZB}, we can compute the the relative phase \eqref{eq:tachyonmassA} 
at any point on the real axis in the $\alpha$-plane except the conifold point 
$\alpha=4$. The conifold point can be avoided following a circular contour of 
very small radius $\epsilon$ centered at $\alpha=4$. 

\begin{figure}[ht]
\begin{center}
\includegraphics[scale=0.5]{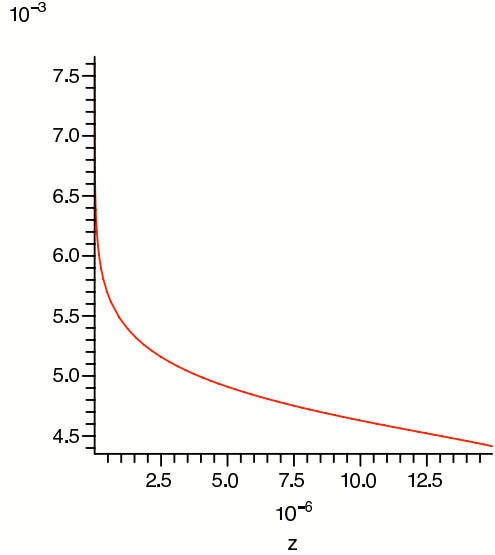} 
\end{center}
\caption{The behavior of the relative phase $\Delta\phi$ in the geometric
  phase for $p=10$.}
\label{shapeB}
\end{figure}

\noindent
The graph in fig. 2 represents the dependence 
of $\Delta \phi$ as a function of $z=\alpha^{-8}$ in the large radius 
phase $0< z < 4$ for $p=10$. Note that it decreases monotonically from 
0.0075 to 0.0044 as we approach the conifold point. Using formulas
\eqref{eq:exactZB}, 
we find that in the LG phase $0<\alpha <4$, $\Delta \Phi$ also decreases 
monotonically until it reaches the value $0.027$ at the LG point. 
One can also calculate the values of
$\Delta\phi$ along a small circular contour surrounding the conifold, 
confirming that it varies continuously in this region. Since $\Delta \phi <1$,
everywhere on the real axis, we conclude that the magnetized brane 
${\underline {\CO_C}}(q)$ is stable with respect to the decay channel 
\eqref{eq:decayA}. 

The analysis of other decay channels is very similar. Another decay channel 
is given by the following short exact sequence 
\be\label{eq:decayB} 
0 \to \CO_D(-C)(q)\to \CO_D(q) \to \CO_C(q) \to 0
\ee
where $D$ is a divisor on $X$ in the linear system $\eta_1$ 
containing $C$. 
Then, an analogous computation yields a similar variation of 
$\Delta\phi$ on the real axis, except that the maximum value is
approximatively $0.015$ and it decreases monotonically 
to $0.008$ at the LG point. 
Therefore the magnetized brane is also stable with respect to the decay 
\eqref{eq:decayB}. In principle there could exit other decay channels, perhaps 
described by more exotic triangles in the derived category. A systematic 
analysis would take us too far afield, so we will simply assume that the 
magnetized branes are stable at the LG point based on the evidence presented 
so far. A rigorous proof of stability is not within the reach of current 
$\Pi$-stability techniques.

\bibliography{meta}
\bibliographystyle{utcaps}

\end{document}